\documentclass[%
  aps,
  prx,              
  twocolumn,
  superscriptaddress,
  10pt
]{revtex4-1}

\usepackage{braket}
\usepackage{graphicx}
\usepackage[colorlinks=true,linkcolor=blue,citecolor=blue,urlcolor=blue]{hyperref}
\usepackage{amsmath}
\usepackage{float}

\begin{document}

\title{Experimental observation of dynamical blockade between\\ transmon qubits via ZZ interaction engineering}

\newcommand{\affPlanckian}{Planckian, I-56127 Pisa, Italy}
\newcommand{\affNapoliPhys}{Dipartimento di Fisica, Universit\`a di Napoli ``Federico II'', Via Claudio, I-80125 Napoli, Italy}
\newcommand{\affNEST}{NEST, Scuola Normale Superiore, Piazza dei Cavalieri 7, I-56127 Pisa, Italy}
\newcommand{\affNapoliIng}{Dipartimento di Ingegneria Elettrica e delle Tecnologie per l'Informazione, Universit\`a di Napoli ``Federico II'', Via Claudio, I-80126 Napoli, Italy}
\newcommand{\affPisaPhys}{Dipartimento di Fisica, Universit\`a di Pisa, Largo Bruno Pontecorvo 3, I-56127 Pisa, Italy}

\author{M.~Riccardi}
\email{mriccardi@planckian.co}
\thanks{These authors contributed equally to this work.}
\affiliation{\affPlanckian}

\author{A.~Glezer~Moshe}
\thanks{These authors contributed equally to this work.}
\affiliation{\affPlanckian}

\author{G.~Menichetti}
\affiliation{\affPlanckian}

\author{R.~Aiudi}
\affiliation{\affPlanckian}

\author{C.~Cosenza}
\affiliation{\affNapoliPhys}

\author{A.~Abedi}
\affiliation{\affPlanckian}

\author{R.~Menta}
\affiliation{\affPlanckian}
\affiliation{\affNEST}

\author{H.~G.~Ahmad}
\affiliation{\affNapoliPhys}

\author{D.~Nieri~Orfatti}
\affiliation{\affPlanckian}

\author{F.~Cioni}
\affiliation{\affNEST}

\author{D.~Massarotti}
\affiliation{\affNapoliIng}

\author{F.~Tafuri}
\affiliation{\affNapoliPhys}

\author{V.~Giovannetti}
\affiliation{\affPlanckian}
\affiliation{\affNEST}

\author{M.~Polini}
\affiliation{\affPlanckian}
\affiliation{\affPisaPhys}

\author{F.~Caravelli}
\affiliation{\affPlanckian}

\author{D.~Szombati}
\email{dszombati@planckian.co}
\affiliation{\affPlanckian}


\begin{abstract}
We report the experimental realization of strong longitudinal (ZZ) coupling between two superconducting transmon qubits achieved solely through capacitive engineering. By systematically varying the qubit frequency detuning, we measure cross-Kerr inter-qubit interaction strengths ranging from $10~\text{MHz}$ up to $350~\text{MHz}$, more than an order of magnitude larger than previously observed in similar capacitively coupled systems. In this configuration, the qubits enter a strong-interaction regime in which the excitation of one qubit inhibits that of its neighbor, demonstrating a dynamical blockade mediated entirely by the engineered ZZ coupling. Circuit quantization simulations accurately reproduce the experimental results, while perturbative models confirm the theoretical origin of the energy shift as a hybridization between the computational states and higher-excitation manifolds. We establish a robust and scalable method to access interaction-dominated physics in superconducting circuits, providing a pathway towards solid-state implementations of globally controlled quantum architectures and cooperative many-body dynamics.
\end{abstract}

\maketitle

\section*{Introduction}
The controlled engineering of interactions between quantum systems underpins experimental quantum science and quantum information processing.  The form, symmetry, and strength of these interactions determine the accessible many-body Hamiltonians and thus constrain both the physical phenomena that can be explored and the computational capabilities of a given platform \cite{Feynman1982, Lloyd1996, NielsenChuang}. Across both analog and gate-based quantum processors, a key challenge is to realize interactions that are simultaneously strong, customizable, and that can be easily addressed as the system size increases \cite{Georgescu2014, Preskill2018}. These are usually conflicting requirements, as exemplified by superconducting quantum architectures \cite{DevoretSchoelkopf2013, Wendin2017, Krantz2019}. Here, scalability is limited not only by decoherence and control/readout fidelity, but also by the physical complexity of the control hardware. In conventional layouts, the requirement for individual microwave control for each qubit leads to a rapidly growing wiring and input/output bottleneck as the device size increases, motivating new approaches to control sharing \cite{Zhao2024RowColumn}, cryogenic electronics \cite{Mukhanov2011, Charbon2022CryoCMOS, Boyer2022CryoMux, Howe2022, Liu2023, Shi2023MultiplexedControl}, and three-dimensional integration \cite{Rosenberg2017, Brecht2015Multilayer}. Despite these limitations, superconducting transmon qubits~\cite{DevoretSchoelkopf2013, Wendin2017, Krantz2019, Koch2007, YouNori2005, SchoelkopfGirvin2008, Kjaergaard2020, Bravyi2022, Ezratty2023} remain a leading platform for near-term scaling, largely because their circuit quantum electrodynamics (cQED)~\cite{Blais2021} architecture allows great flexibility in tuning the interactions between different quantum systems and because their fabrication workflows are relatively mature and reproducible~\cite{DevoretSchoelkopf2013,Krantz2019, Kjaergaard2020,Bravyi2022}. 
As a result, there are already several commercial multi-qubit programmable superconducting processors, which have led to the demonstration of quantum computational advantage~\cite{Arute2019, Wu2021, Kim2023}, phase-sensitive two-qubit gates~\cite{CK2011, Sheldon2016}, and the simulation of cooperative phenomena in fully artificial solid-state quantum systems
~\cite{GagoEncinas2023Graph,GagoEncinas2023DimExp,GagoEncinas2025Modular}.

On the other hand, in atomic and molecular systems, the available couplings are set by microscopic forces and can give rise to striking collective phenomena. Neutral atoms excited to high-lying Rydberg states provide a particularly powerful example. Their large dipole--dipole and Van der Waals interactions allow for fast multi-qubit gates, strong blockade effects, and the preparation of highly correlated many-body states, paving the way for both quantum gate proposals and large-scale quantum simulators \cite{Jaksch2000, Saffman2010, BrowaeysLahaye2020, Bernien2017, Gaetan2009, Urban2009}.

Building on these capabilities, Rydberg arrays have inspired the design of globally controlled quantum architectures in which large ensembles of qubits are manipulated through shared control fields rather than individually routed signals \cite{Cesa2023}, providing a clear route towards scalability by reducing the number of required control signals required to execute quantum computation. However, the limited in-situ tunability of these interactions motivates transposing the underlying ideas to engineered quantum platforms that offer greater flexibility, suggesting a pathway in which a Rydberg-inspired blockade generated by large longitudinal (ZZ) interactions is implemented in superconducting circuits to realize globally controlled architectures~\cite{Menta2025, Cioni2024, Menta2025b}.

However, in most transmon architectures, the inherent longitudinal interaction is relatively weak, so sizable effective ZZ terms typically arise only indirectly (e.g. via higher-order processes, residual coupler-mediated interactions, or gate-induced dressing) and are typically regarded as unwanted spurious effects which contribute to crosstalk and residual two-qubit gate errors. More specifically, the ZZ interaction between qubits has been studied by coupling the qubits via a bus resonator~\cite{DiCarlo2009, Chow2011, McKay2019, Huang2024} or a tunable coupler~\cite{Kounalakis2018, Collodo2020, Ganzhorn2020, Xu2020, Cai2021, Chu2021, Sung2021, Ni2022} in order to either suppress this interaction and reduce unwanted phase accumulation during quantum computation \cite{McKay2019, Lacroix2020}, or enhance its on/off ratio to execute fast CZ gates \cite{Collodo2020, Sung2021, Marxer2023, Li2024}. Coupling qubits capacitively also yields a ZZ interaction \cite{Caldwell2018, Karamlou2021, Long2021, Wei2022} and is easier to target as it does not require the precise fabrication of extra junctions with a specified Josephson energy.
By employing this coupling scheme, to the best of our knowledge, some of the largest ZZ couplings measured have been 9.29 MHz for a qubit detuning of 240 MHz \cite{Long2021} and 4 MHz for a detuning of 642 MHz \cite{Caldwell2018}. However, in order to realize superconducting architectures for global quantum computing, the ZZ interaction strength $\zeta$ in these systems needs to be the dominant energy scale with respect to the qubit's Rabi drive frequency $\Omega$ \cite{Menta2025, Cioni2024, Menta2025b}, which usually has a magnitude of a few tens of MHz. This guaranties that the inequality $\Omega \ll \alpha$ is satisfied, where $\alpha \sim 300~{\rm MHz}$ is the qubit anharmonicity, which is the standard requirement for transmon qubits to remain in the computational subspace during their manipulations. Our goal is therefore to look for $\zeta$ of a few hundreds of MHz while maintaining a substantial detuning $\Delta$ between qubits in order to avoid drive crosstalk. Engineering such nearest neighbor coupling is not trivial, as qubit-qubit interactions generally scale with $1/\Delta$.

In this work, we present a detailed and thorough investigation of the ZZ interaction $\zeta$ between two capacitively coupled transmon qubits using spectroscopic measurements. By varying the frequency detuning between the qubits, we observe interaction strengths ranging from 10 MHz at a detuning of 2 GHz up to an exceptionally strong 350 MHz when the qubits are set near to resonance. We confirm our findings by reproducing similar data from a second device, where we measured a $\zeta$ of similar magnitude. This longitudinal coupling is more than an order of magnitude greater than any previously reported value in the literature for this experimental platform, largely surpassing both typical Rabi drive frequencies and anharmonicities of transmons~\cite{large_ZZ_resonator}. Our experimental data show excellent agreement with theoretical calculations, even at small detunings. We further demonstrate that this enhanced coupling strength is sufficient to place the system in a dynamical blockade regime~\cite{Menta2025} akin to the Rydberg one observed in atomic systems, providing the first observation of this phenomenon in a superconducting platform. Our work paves the way for implementing collective quantum operations and simulating many-body Hamiltonians with long-range correlations in superconducting circuits. 

\section*{ Experimental results}
The experimental devices consist of two capacitively coupled, flux-tunable transmon qubits~\cite{Koch2007}.

\begin{figure*}
    \centering
\includegraphics[width=1.0\linewidth]{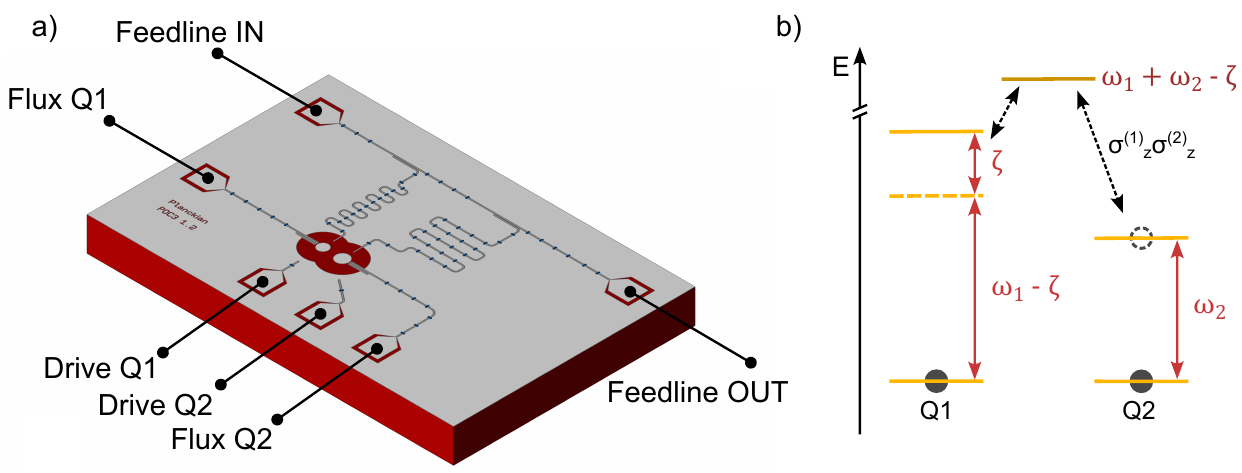}
    \caption{\textbf{(a)} Device under test, composed of two transmon-type qubits (Q1 and Q2) with mutual capacitive coupling. \textbf{(b)} Effect of the ZZ interaction. When Q1 and Q2 are in the ground state, they resonate at $\omega_{1,2}$ respectively (full lines). The excitation of one qubit (Q2) leads to a shift of the energy levels of the adjacent qubit (Q1) by $\zeta$ (dashed lines).}
    \label{fig:fig1}
\end{figure*}
 For the first device, which is presented in the main text, each qubit consists of a superconducting quantum interference device (SQUID) that shunts a disk-shaped capacitor to the ground plane, forming a compact and low-loss circuit element (see Figure~\ref{fig:fig1}a). The qubits, labeled Q1 and Q2, are positioned in close proximity to realize a direct capacitive coupling, while retaining independent frequency control via dedicated flux bias lines. Individual microwave drive lines enable selective single-qubit manipulation, and each qubit is dispersively coupled to its own $\lambda/4$ coplanar-waveguide readout resonator, allowing independent state measurement through heterodyne detection. 

The transition frequencies $\omega_{1,2}/(2\pi)$ of Q1 and Q2 can be continuously tuned via the flux bias lines from $6.270~\text{GHz}$ to  $ 9.197~\text{GHz}$ for Q1 and from $4.200~\text{GHz}$ to $6.348~\text{GHz}$ for Q2, while the corresponding resonator frequencies are $\omega_{r1}/(2\pi) = 7.350~\text{GHz}$ and $\omega_{r2}/(2\pi) = 4.882~\text{GHz}$ when each qubit is tuned to its maximum frequency. 
The measured anharmonicities are $\alpha_{1}/(2\pi) \approx -351~\text{MHz}$ and $\alpha_{2}/(2\pi) \approx -312~\text{MHz}$, consistent with the expected transmon regime~\cite{Koch2007}. 
The capacitive coupling element has a design value of $C_{12} = 5.11~\text{fF}$, corresponding to an estimated exchange interaction strength of $J/(2\pi) \approx 240~\text{MHz}$, as predicted by the numerical diagonalization of the circuit Hamiltonian and confirmed by spectroscopic measurements (see the Supplementary Information (SI)~\cite{SI}).
At its frequency minimum, Q1 (Q2) exhibits an average $T_1$ energy-relaxation time of $7.0~\mu\text{s}$ ($7.8~\mu\text{s}$) and $T_2^{\text{echo}}$ echo coherence times of $9.0~\mu\text{s}$ ($6.3~\mu\text{s}$). 
A summary of all relevant device parameters, including capacitances, frequencies, and coherence metrics, is provided in Table~I of the SI~\cite{SI}.

We also show in the SI~\cite{SI} the results of measurements performed on a second device, with the qubit capacitor shaped in the X-mon geometry and different coherence properties, but nominally the same circuit parameters. For this device, we successfully reproduce all the results presented in the main text for the first chip, providing a solid experimental validation of the physics studied in this work.

\subsection*{ZZ coupling as a function of qubit detuning}
The direct capacitance between Q1 and Q2 results in a longitudinal interaction that can be modeled via an effective two-qubit interaction Hamiltonian $H_{\rm int}$ arising perturbatively from the nonlinear Josephson terms and capacitive coupling between the qubits~\cite{Kounalakis2018,Solgun2022}, and can be written as
\begin{equation}
    H_{\text{int}} = \frac{\zeta(\Delta)}{4}\, \sigma_{z}^{(1)} \sigma_{z}^{(2)},
\end{equation}
where $\sigma^{(1,2)}_z$ represent Q1 and Q2 as effective two-level systems and $\Delta= \omega_{1}-\omega_{2}$ is the bare detuning between the qubits defined when $H_{\text{int}}=0$. This Hamiltonian describes a cross-Kerr interaction, i.e.~a shift of magnitude $\zeta$ in the transition frequency of each qubit, depending on the state of the other qubit, as shown in the energy level diagram depicted in Figure \ref{fig:fig1}b.

\begin{figure*}[t!]
\centering
\includegraphics[width=1.0\textwidth]{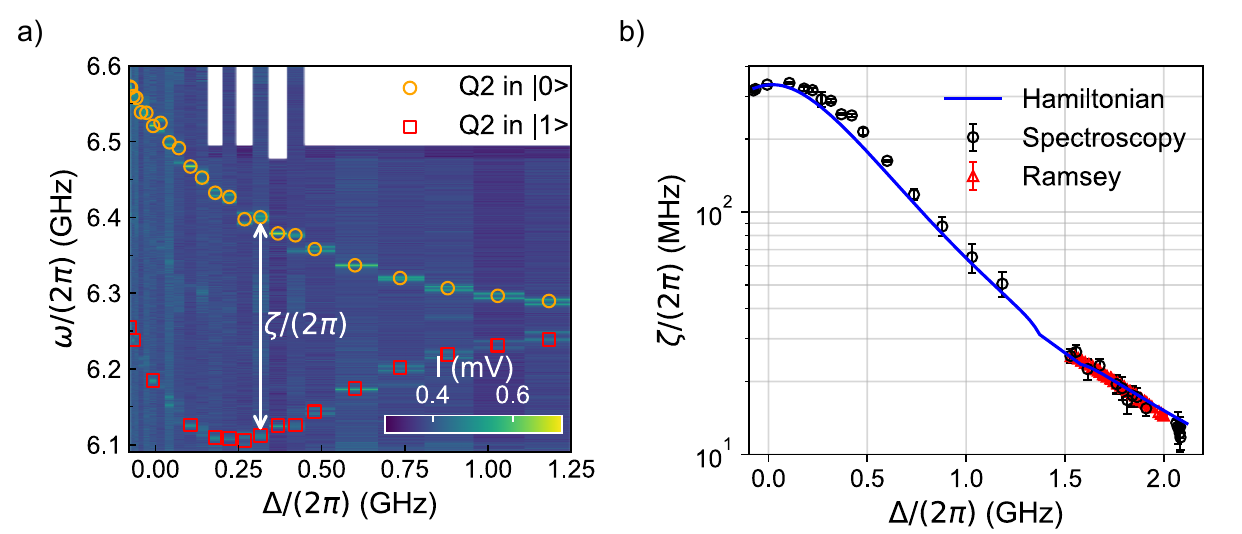}
\caption{\textbf{(a)} Q1 spectroscopy as a function of $\Delta / (2 \pi)$ when Q2 is either in the ground or excited state. The orange and red markers show the extracted Q1 frequency from the spectroscopy with a Lorentzian fit when Q2 is in $\ket{0}$ or $\ket{1}$ respectively. $\zeta$ is computed as the difference between these two frequencies. \textbf{(b)} Measured ZZ interaction as a function of $\Delta / (2 \pi)$. The black circles mark the ZZ extracted from the spectroscopy measurements shown in panel (a), while the red triangles are extracted from Ramsey measurements on Q1 conditional on the state of Q2. Additionally, we show in blue the ZZ extracted from the circuit Hamiltonian. The kink in the theoretical curve around $\Delta / (2 \pi) = 1.4$ GHz is caused by the readout resonator of Q2 located at that frequency.}
\label{fig:ZZvsDelta}
\end{figure*}

We experimentally characterize $\zeta(\Delta)$ by performing a two-tone spectroscopy on Q1 while varying the frequency of Q2, and hence $\Delta$, via its own flux-bias line. With Q1 parked at its lower sweet-spot, the detuning between the qubits was varied from $2.0~{\rm GHz}$ down to the resonant regime. At each value of $\Delta$, the frequency of Q1 was measured twice: once with Q2 in the $|0\rangle$ state and once after preparing Q2 in $|1\rangle$ via a $\pi$-pulse. Figure~\ref{fig:ZZvsDelta}a shows the results of these spectroscopy measurements, where several datasets are superimposed to reveal the frequency of Q1 in both scenarios. For each value of $\Delta$, we marked with an orange circle the fitted spectroscopy peak acquired when Q2 is measured in $|0\rangle$ and  with a red square when Q2 is measured in $|1\rangle$. As depicted in Figure ~\ref{fig:fig1}b, these peaks represent the $\omega_1$ and $\omega_1 - \zeta$ transition respectively, from which the interaction strength $\zeta$ can be directly extracted as the energy difference between these two levels.

The full dependence of  $\zeta$ on $\Delta$ is presented in Figure ~\ref{fig:ZZvsDelta}b. The measured interaction strength spans nearly two orders of magnitude across the accessible range of detunings, varying from $\zeta/(2\pi) = 10~\text{MHz}$ at $\Delta/(2\pi) \approx 2~\text{ GHz}$ up to an exceptionally strong $\zeta/(2\pi) = 350~\text{ MHz}$ at $\Delta \approx 0$. 
This represents, to our knowledge, one of the largest capacitive ZZ couplings reported~\cite{Caldwell2018, Long2021}. 
The monotonic increase of $\zeta$ with decreasing $\Delta$ is consistent with the expected $\propto J^{2}/\Delta$ scaling of the effective cross-Kerr interaction~\cite{Solgun2022}. More importantly, we find that this strong interaction remains well-described by standard circuit Hamiltonian models even as the system approaches the non-perturbative regime, where $J/\Delta$ is no longer small. 

To further validate the extracted $\zeta$ values, we performed an independent cross-check via time-domain measurements by measuring $\omega_1$ via Ramsey interferometry, similar to what was done in \cite{DiCarlo2009}, when Q2 is in either $|0\rangle$ or $|1\rangle$. The spectroscopy and time-domain methods are in excellent agreement, confirming the reliability of the extracted interaction strengths. We note that due to the limited time resolution ($4~{\rm ns}$) of our electronics  between each point, the time-domain method to measure detuning is limited to a value of $\zeta/(2\pi) < 125\text{ MHz}$ due to aliasing at higher frequencies.

Together, these measurements demonstrate in-situ tunability of the longitudinal coupling over a wide range, achieved simply through flux control of one qubit, and establish a direct experimental route to achieve the strong-interaction regime required for realizing blockade phenomena and collective dynamics in superconducting qubit arrays.

\subsection*{Blockade experiment}

The strong ZZ coupling measured in our devices enables the realization of a dynamical blockade regime in a cQED  architecture~\cite{Menta2025}, akin to that occurring in Rydberg atoms~ \cite{Gaetan2009, Urban2009}. 

\begin{figure}
\includegraphics[width=1.0\columnwidth]{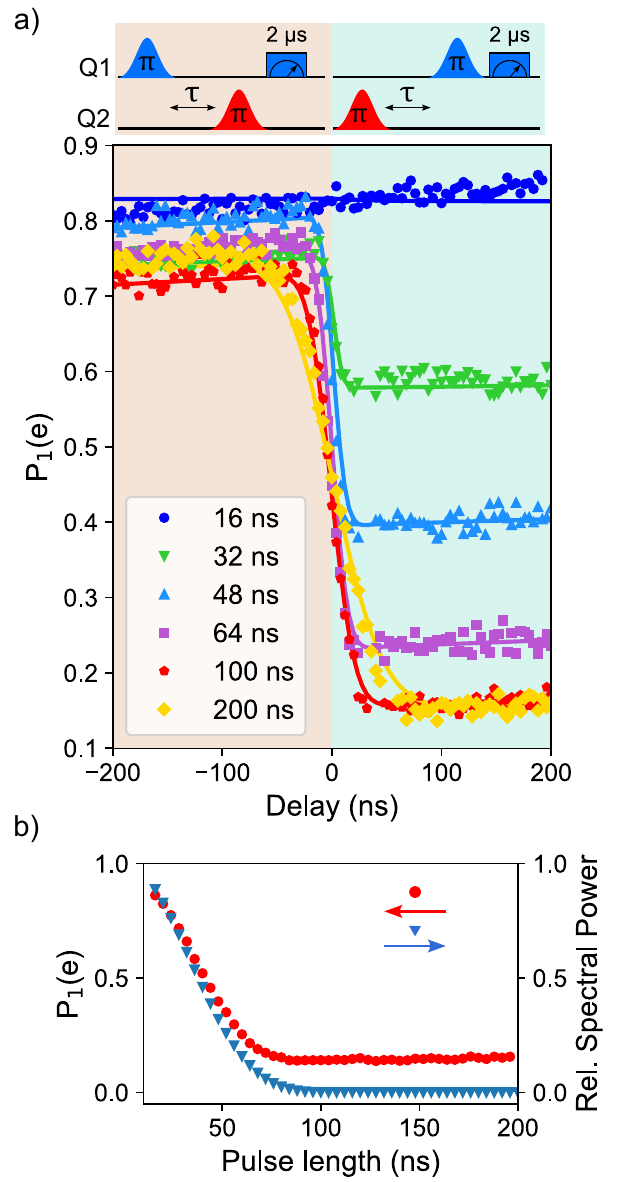}
\caption{(\textbf{a}) Measured (points) and predicted (full lines) populations of Q1 versus relative delay for six different excitation pulse lengths ($16$ ns - $200$ ns). When long pulses are applied, exciting Q2 first (positive delay) inhibits the excitation of Q1, demonstrating blockade due to the qubit cross-Kerr coupling ($\zeta/(2\pi) = 19~\text{MHz}$). Top inset: pulse sequences for Q1 and Q2, illustrating positive and negative excitation delays.
(\textbf{b}) The red points show the evolution of $P_1(e)$ as a function of the excitation pulse length for a delay of 100 ns. The blue points show, for each pulse length, the relative power of the drive pulse integrated over a spectral window centered around $(\omega_1 - \zeta) / (2 \pi)$ and with a width given by $1 / T_2^{Q1}$.}

\label{fig:blockade}

\end{figure}

We demonstrate this by consecutively exciting the two qubits with a pulse of varying length and a truncated cosine envelope when the qubits are at $\omega_1/(2\pi) = 6.307$ GHz and $\omega_2/(2\pi) = 4.498$ GHz, where $\zeta = 19$ MHz. The excited state population of Q1, $P_1(e)$, is then measured as a function of the time delay between the qubit excitations, as shown in Figure~\ref{fig:blockade}a. For negative time delays, Q1 is excited prior to Q2 and therefore $P_1(e)$ reaches the level expected from the $\pi$-pulse and readout fidelity (which is slightly higher for shorter pulses). This behavior  changes drastically for positive delay times (turquoise-shaded area in Figure~\ref{fig:blockade}a) where Q1 is excited following the excitation of Q2. If the $\pi$-pulse  is long, Q1 remains mostly in the ground state due to the blockade effect, limited only by the excitation efficiency of Q2. For each curve, the width of the crossover in $P_1(e)$ around 0 delay is comparable to the drive pulse length.
As the length of the $\pi$-pulse on Q1 shortens, $P_1(e)$ increases and the blockade becomes less effective, as the pulse carries a larger frequency component corresponding to the blockade transition $\omega_1-\zeta$. 

To further investigate the effect of the $\pi$-pulse length on the blockade, in Figure \ref{fig:blockade}b we plot $P_1(e)$ for a fixed delay of $100$ ns as a function of the pulse length, together with the relative power of the pulse at frequency  $\omega_1-\zeta$. As expected, the rise of $P_1(e)$ correlates well with the increase in the $\pi$-pulse's frequency component matching $\omega_1-\zeta$.

For long time delays on the order of microseconds, the blockade effect vanishes at an exponential rate, matching well the $T_1$ decay time of Q2, which relaxes during the delay allowing Q1 to be excited by a $\pi$-pulse resonant with $\omega_1^g$ (see the plot in the SI~\cite{SI}).

\section*{Theoretical analysis}
From a theoretical point of view, the longitudinal interaction between two capacitively coupled transmon qubits originates from the weak Josephson anharmonicity of each mode together with their mutual capacitive hybridization. 
In the dispersive regime, where the detuning $\Delta=\omega_1-\omega_2$ is large compared to the exchange rate, the coupled system can be modeled as two weakly anharmonic oscillators with an exchange interaction,
\begin{equation}
    \hat{H}
    =
    \sum_{i=1}^{2}
    \left(
        \omega_i \hat{a}_i^\dagger \hat{a}_i
        +\frac{\alpha_i}{2}\hat{a}_i^{\dagger 2}\hat{a}_i^{2}
    \right)
    + g\left(\hat{a}_1^\dagger \hat{a}_2+\hat{a}_1\hat{a}_2^\dagger\right),
    \label{eq:H_bare_main}
\end{equation}
where $\hat{a}_i$ are bosonic annihilation operators, $\alpha_i$ are the transmon anharmonicities and $g\simeq J$ is the capacitive exchange coupling~\cite{Koch2007,Blais2004}. 
Although Eq.~\eqref{eq:H_bare_main} contains no explicit $\hat{n}_1\hat{n}_2$ term, an effective ZZ interaction is generated perturbatively by virtual excursions from the computational manifold into the two-excitation non-computational states $\ket{20}$ and $\ket{02}$.

As detailed in the SI, the emergence of the ZZ term is captured by truncating each oscillator to the lowest three levels and performing a Schrieffer--Wolff elimination of the non-computational two-excitation subspace to second order in $g$.
This yields a dressed energy shift of $\ket{11}$ relative to $\ket{10}+\ket{01}$, quantified by
\begin{equation}
\zeta
=
E_{11}-E_{10}-E_{01}+E_{00},
\label{eq:zeta_def_main}
\end{equation}
and a corresponding effective qubit Hamiltonian
\begin{eqnarray}
\hat{H}_{\mathrm{eff}}
&=&
\frac{\tilde{\omega}_1}{2}\sigma_z^{(1)}
+
\frac{\tilde{\omega}_2}{2}\sigma_z^{(2)}\nonumber \\
&&+
\frac{J}{2}\!\left(\sigma_x^{(1)}\sigma_x^{(2)}+\sigma_y^{(1)}\sigma_y^{(2)}\right)
+
\frac{\zeta}{4}\,\sigma_z^{(1)}\sigma_z^{(2)}\nonumber \\
&&+\mathrm{const.},
\label{eq:Heff_main}
\end{eqnarray}
where $\tilde{\omega}_{1,2}$ include the Lamb shifts.
In the same perturbative limit, the ZZ coefficient can be written as \cite{DiCarlo2009, SI} 
\begin{equation}
\zeta
\simeq
2g^2
\left(
\frac{1}{\Delta + |\alpha_2|}
-
\frac{1}{\Delta - |\alpha_1|}
\right),
\label{eq:zeta_SW_main}
\end{equation}
which makes it explicit that at high detuning the ZZ coupling follows $\zeta\propto g^2/\Delta$ and it is mediated by virtual mixing with $\ket{20}$ and $\ket{02}$.

A useful equivalent representation follows from rewriting the ZZ term in the computational basis. Up to one-body Z terms and an overall energy shift, one may write
\begin{equation}
\frac{\zeta}{4}\sigma_z^{(1)}\sigma_z^{(2)}
\equiv
\zeta\,\ket{11}\!\bra{11}
\;+\;\text{(one-body terms)}.
\label{eq:proj_main}
\end{equation}
This projector form highlights the physical interpretation: $\zeta$ is an interaction energy cost associated with simultaneous excitation of both qubits, directly analogous (at the two-site level) to the blockade term that naturally appears in Rydberg-atom Hamiltonians. 
In particular, in the rotating frame used for our pulse sequences \cite{SI}, choosing the drive frequencies to cancel the residual single-qubit Z contributions yields an effective driven model
\begin{equation}
\hat{H}_{\mathrm{block}}
=
\zeta\,\ket{11}\!\bra{11}
+
\sum_{i=1}^{2}\frac{\Omega_i(t)}{2}\,\sigma_x^{(i)},
\label{eq:H_block_main}
\end{equation}
which shows how a sufficiently large $\zeta$ inhibits the excitation of a qubit when its neighbor is already in the $\ket{1}$ state.

To validate the microscopic origin of the longitudinal interaction beyond perturbation theory, we compare the measured $\zeta(\Delta)$ with a numerical model. In particular, we compute $\zeta(\Delta)$ numerically from the full circuit Hamiltonian using black-box quantization \cite{SI}. 
Starting from the measured or simulated junction admittances, we construct an equivalent Foster network, quantize the resulting multi-mode linear environment, and reintroduce the Josephson nonlinearities perturbatively to obtain the Kerr Hamiltonian parameters. From the dressed spectrum of the truncated Hamiltonian we then evaluate $\zeta$ directly via Eq.~\eqref{eq:zeta_def_main}. 
This procedure is implemented using \texttt{QuCAT} \cite{SI} and provides an estimate of $\zeta$ across the explored detuning range. We use the same method to fit the measured excitation spectra of the two qubits, showing the deviation from the bare qubit spectra as a result of the exchange interaction (see the SI~\cite{SI}).

Finally, we also connect the extracted interaction strength to the observed blockade dynamics. 
We perform a two-qubit blockade pulse sequence in which the relative delay between two $\pi$-pulses ranges from -40 ${\rm ns}$ to 40 ${\rm ns}$, in order to explore the blockade effect on both qubits (see SI~\cite{SI}, Fig.S14 and Fig.S15). 
When Q2 is excited first (positive delay), the subsequent excitation of Q1 is suppressed, consistent with a blockade induced by the longitudinal interaction energy penalty on $\ket{11}$. 
Over longer delays, the population dynamics exhibit exponential relaxation, consistent with the independently measured $T_1$ values (see SI, Fig.~S11), indicating that the blockade signal is not an artifact of slow drift or readout effects. 
To further confirm the mechanism, we simulate the driven two-qubit dynamics using the effective rotating-frame Hamiltonian \eqref{eq:H_block_main} (SI), reproducing the qualitative suppression of excitation and its dependence on delay and pulse parameters.
Altogether, the spectroscopy-based extraction of $\zeta$, the agreement with both perturbative and black-box quantization models, and the time-domain blockade dynamics provide a consistent quantitative picture of the engineered dynamical blockade mediated by strong ZZ coupling in this device.

\section*{Discussion}

The observation of a dynamical blockade between two superconducting qubits establishes that interaction-dominated physics, long associated with atomic platforms, can be realized in a solid-state circuit. By engineering a strong longitudinal (ZZ) coupling through mutual capacitance alone, we show that a single excitation can shift a neighboring qubit out of resonance and thereby inhibit simultaneous excitations. This reproduces the essential mechanism of a dipole blockade in Rydberg atoms, whereby strong state-dependent interactions constrain the accessible Hilbert space and give rise to cooperative behavior~\cite{Jaksch2000,Saffman2010, BrowaeysLahaye2020, Bernien2017}. 

This result carries significant implications on future quantum computing architectures based on superconducting circuits, where residual longitudinal interactions are usually treated as a small, unwanted byproduct, which must be minimized to protect gate fidelity and reduce coherent crosstalk~\cite{Krantz2019, Bravyi2022, Sheldon2016, Chow2011}. Here, instead, the ZZ coupling is intentionally made large enough to become the dominant energy scale of the driven dynamics, turning what is usually a parasitic effect into a resource for implementing constrained dynamics and interaction-enabled global quantum computing~\cite{Menta2025}.

The blockade efficiency is not determined by the interaction strength alone, but is also limited by control and measurement errors. In particular, the protocol highlights two practical requirements. First, the preparation of the blockading excitation must be sufficiently faithful. Any reduction in the excited-state population of the control qubit directly reduces the observed suppression. Second, the drive applied to the blockaded qubit must be spectrally narrow compared with the blockade-induced shift. If the target $\pi$-pulse is too short, its broad spectrum can overlap with both the inherent and shifted transition frequencies, partially restoring excitation even in the blockaded configuration. These constraints naturally motivate the use of longer, better-shaped control pulses and systematic calibration strategies that reduce leakage, suppress unwanted frequency components, and mitigate drive-induced errors~\cite{Krantz2019, Bravyi2022, Blais2021}. However, when $|\zeta| > |\alpha|$, the length of the $\pi$-pulse is still limited by the anharmonicity alone, as in more standard configurations. Importantly, we have established that superconducting qubits can access this regime for $\Delta <200$ MHz (see Figure \ref{fig:ZZvsDelta}b and S8), demonstrating that a globally controlled architecture can be driven with adequately short pulses while also mitigating any drive crosstalk between neighboring qubits.
 
Extending the pairwise blockade demonstrated here to chains or lattices would enable collective blockade phenomena, where one (or a few) excitations inhibit dynamics across extended qubit networks and where global driving fields can generate correlated dynamics under constrained motion~\cite{Saffman2010, BrowaeysLahaye2020, Bernien2017}. Such ``blockade networks'' would offer a solid-state route to regimes that have been extensively explored in neutral-atom arrays, while benefiting from the fast control, mature fabrication, and tight integration available in superconducting hardware~\cite{Kjaergaard2020, Bravyi2022, Blais2021}.

Finally, strong, tunable ZZ interactions broaden the toolbox for superconducting quantum simulation and interaction-driven programming. They enable direct implementation of constrained models and interaction-enabled protocols in a platform traditionally optimized for gate-based operation, complementing existing approaches to quantum simulation in the NISQ (Noisy and Intermediate Scale Quantum) era~\cite{Feynman1982, Lloyd1996, Georgescu2014, Preskill2018}. Crucially, this approach enables strong-interaction regimes without increasing the number of control lines, thereby addressing one of the primary hardware bottlenecks in QPU scalability. Together, these results position capacitive longitudinal coupling as a practical route to realizing cooperative blockade physics and scaling interaction-dominated superconducting quantum circuits. In particular, this is an important step towards the implementation of globally controlled quantum architectures based on ZZ interactions \cite{Menta2025, Cioni2024, Menta2025b, Aiudi2025Disorder}.

\section*{Methods}

\subsection*{Experimental methods}
The sample layouts for both devices were designed using Qiskit Metal, and their electromagnetic properties were simulated with Ansys HFSS and Q3D to set the target qubit frequencies, couplings, and resonator dispersive shifts.

The transmon qubits were fabricated by ConScience on high-resistivity silicon substrates, with aluminum forming the ground plane, resonators, and capacitor pads. Airbridges are used at regular intervals  to connect the ground plane across each co-planar waveguide on the chip.

We study two planar superconducting devices, each comprising a pair of capacitively coupled, flux-tunable transmon qubits with nominally the same design parameters. Chip~1, detailed in the main text, hosts qubits Q1–Q2 implemented as circular-pad transmons. Q1 is operated mainly near its flux-insensitive lower sweet-spot with transition frequency $\omega_1/(2\pi) \approx 6.3~\text{GHz}$, while Q2 was studied between $\omega_2/(2\pi) \approx 4.2~\text{GHz}$ and $\omega_2/(2\pi) \approx 6.3~\text{GHz}$. The measured anharmonicities were in the order of $-300~\text{MHz}$, and representative coherence times were $T_1 \simeq 8$–$9~\mu$s and $T_2 \simeq 1$–$5~\mu$s (see Table~S1). Chip~2, displayed in the SI, hosts qubits Q3–Q4 implemented in an X-mon geometry, with lower sweet-spot frequencies $\omega_3/(2\pi) \approx 7.9~\text{GHz}$ and $\omega_4/(2\pi) \approx 5.6~\text{GHz}$. This second device is used to corroborate the blockade physics through independent measurements of single-shot IQ clouds and readout fidelities. Chip~1 was mounted in a Quantum Machines QCage.64 sample holder, while Chip~2 was bonded to a custom sample holder. Both samples were cooled in a cryogen-free dilution refrigerator with a base temperature of $\sim 10$~mK.

In both devices, the qubits are coupled to each other via direct capacitive coupling and to individual coplanar-waveguide readout resonators that share a common feedline for dispersive readout. Flux-bias lines provide individual DC and pulsed flux control, while dedicated microwave drive lines enable single-qubit rotations. All input lines are heavily attenuated and filtered at multiple temperature stages using a combination of low-pass and Eccosorb filters to suppress thermal noise and spurious high-frequency radiation. The readout signal propagates through circulators and a band-pass filter to a cryogenic HEMT amplifier at the 4~K stage, followed by room-temperature amplification and heterodyne detection in either a Quantum Machines (Chip~1) or Qblox (Chip~2) microwave front-end. Qubit frequencies and coherence parameters are obtained from standard $T_1$, Ramsey, and spin-echo measurements. The static ZZ interaction $\zeta$ is extracted from spectroscopy, Ramsey and modified-echo sequences in which one qubit is prepared in $\ket{g}$ or $\ket{e}$ while the partner is interrogated, whereas the coherent exchange rate $J$ is determined from spectroscopy. Dynamical blockade is characterized using pairs of calibrated $\pi$-pulses with variable relative delay on the two qubits and by analyzing the resulting populations and single-shot histograms on both chips using custom Python-based data analysis routines. Details are presented in the SI~\cite{SI}.

\subsection*{Theoretical methods}
We combine analytical and numerical approaches to model the coupled-transmon circuit and its driven dynamics. As a first step, we derive an analytical expression for the static cross-Kerr interaction $\zeta$ using a Schrieffer–Wolff treatment of two weakly anharmonic capacitively coupled oscillators. 

This closed-form $\zeta(\Delta, \alpha_{1,2}, g)$ is used as a fast cost function to optimize circuit parameters (capacitances and Josephson energies) under the transmon-regime and fabrication constraints via a differential-evolution algorithm. Above, $\alpha_i$ are the anharmonicity of the $i$-th transmon qubit (see SI~\cite{SI} for a derivation). 
For the resulting designs, we perform black-box quantization \cite{Nigg2012BlackBox} of the linearized circuits with two independent numerical frameworks, \texttt{SQcircuit} and \texttt{QuCAT}~\cite{Rajabzadeh2023analysisofarbitrary,GelySteele2020QuCAT}, obtaining a multimode Kerr Hamiltonian directly from the circuit layout (see SI ~\cite{SI}). From this Hamiltonian, we extract refined values for the qubit frequencies, anharmonicities, the exchange coupling $J$ and the cross-Kerr strength $\zeta$, which are then used as fixed parameters in the effective two-qubit Hamiltonian discussed in the main text. Time-dependent driven dynamics, including the dynamical blockade protocol, are simulated by numerically solving the Schr\"odinger equation for this Hamiltonian with the \texttt{QuTiP} library~\cite{Johansson2012QuTiP,Johansson2013QuTiP2}. The theoretical curves shown in the figures are therefore genuine zero-parameter predictions, fully determined by independently calibrated device and pulse parameters. Details are presented in the SI~\cite{SI}.

\bibliography{bibliography}


\begin{acknowledgments}
We thank D.~De Bernardis for comments on the manuscript.

\paragraph*{\bf Funding.}
This research was privately funded. H.G.A., D.M., and F.T. acknowledge support from the PNRR MUR Projects PNRR PE000023-NQSTI and PNRR CN00000013-ICSC.

\paragraph*{\bf Author contributions.}
M.R. and D.N.O. designed the chips. A.G.M., C.C., and H.G.A. performed the experimental measurements. G.M., R.A., A.A., R.M., and F.C. performed theoretical analysis, fits, and parameter optimization. D.M., F.T., V.G., and M.P. supervised the overall effort. F.C. led the theoretical work and D.S. led the experimental work. All authors contributed to the writing of the manuscript.

\paragraph*{\bf Competing interests.}
M.R., A.G.M., G.M., R.A., A.A., R.M., D.N.O., F.C., V.G., M.P., F.C., and D.S. are funded by Planckian srl, which develops quantum computing hardware and globally controlled superconducting architectures. M.P. and V.G. are co-founders of Planckian srl. M.P., V.G., M.R., and R.A. are shareholders of Planckian srl.

\paragraph*{\bf Data and materials availability.}
Data and codes are available from the corresponding authors upon request.
\end{acknowledgments}

\clearpage
\onecolumngrid


\begin{center}
{\bfseries \large Supplementary Information}
\end{center}

\vspace{0.5cm}

\renewcommand{\thefigure}{S\arabic{figure}}
\renewcommand{\thetable}{S\arabic{table}}
\renewcommand{\theequation}{S\arabic{equation}}
\renewcommand{\thepage}{S\arabic{page}}
\setcounter{figure}{0}
\setcounter{table}{0}
\setcounter{equation}{0}
\setcounter{page}{1} 

\section{Device parameters}
Table \ref{tab:qubit_parameters_main} lists the device parameters of Q1 and Q2 discussed in the main text. LSS, g and $\chi$ represent, respectively, the lower sweet-spot, the coupling between one qubit and its own readout resonator and the dispersive shift.

\begin{table}[h]
\caption{\label{tab:qubit_parameters_main}%
System parameters of Chip~1 discussed in the main text. Values in parentheses denote the qubit frequencies at which that parameter is measured.}
\begin{ruledtabular}
\begin{tabular}{rcc}
 & \textrm{Q1} & \textrm{Q2} \\

$\omega_{USS}/(2\pi)$ (GHz)&  9.20&  6.35\\
 $\omega_{LSS}/(2\pi)$ (GHz)& 6.27& 4.20\\
$\alpha/(2\pi)$ (MHz) &  -351&  -312\\
$\omega_r/(2\pi)$ (GHz)&  7.350&  4.882\\
g$/(2\pi)$ (MHz) &  51&  32\\
$\chi/(2\pi)$ (kHz) &  -375 (LSS)&  -1380 (4.5 GHz)\\
T$_1$ ($\mu$s) & 7.8$\pm$0.4 (LSS)&  8.8$\pm$0.4 (4.5 GHz)\\
T$_2^*$ ($\mu$s) &  5.0$\pm$0.2 (LSS)&  1.1$\pm$0.2 (4.5 GHz)\\
T$_2^{echo}$ ($\mu$s)&  9.1$\pm$1.4 (LSS)&  2.3$\pm$0.2 (4.5 GHz)\\
\end{tabular}
\end{ruledtabular}
\end{table}

\newpage

\section{Measurement Setup}
\begin{figure*}[h]
\includegraphics[width=0.7\textwidth]{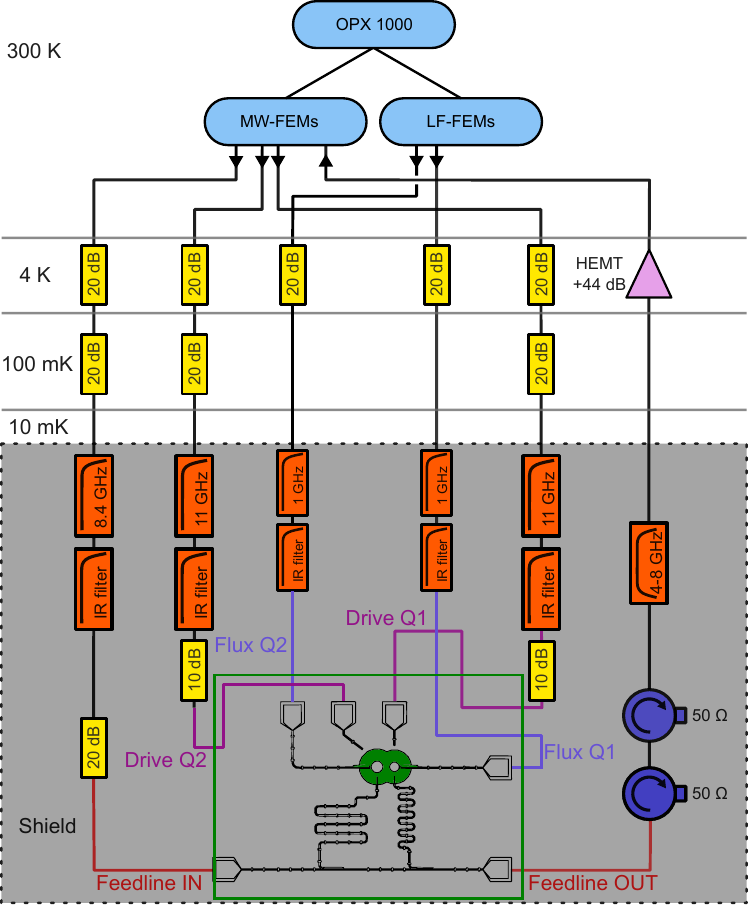}
\caption{\label{fig:Wiring_IQCC} The measurement setup. The sample is marked with green for clarity. The arrows near each FEM marks the direction of the signal flow.}
\end{figure*}

The diagram of the setup inside the fridge used for Chip~1 is shown in Fig. \ref{fig:Wiring_IQCC}.
The sample was mounted on a commercial Quantum Machines QCage.64 holder and enclosed in a shield consisting of an outer $\mu$-metal layer and an inner aluminum surface. Attenuators were placed at several stages, providing 60 dB attenuation on the readout line, 50 dB on the drive lines, and 20 dB on the flux lines. The readout line includes an 8.4 GHz low-pass filter and an Eccosorb (IR) filter, the drive lines use 11 GHz low-pass and Eccosorb filters, and the flux lines use 1 GHz low-pass and Eccosorb filters. The readout signal from a Quantum Machines OPX1000 microwave front-end module (MW-FEM), after passing through the sample feedline, travels through two circulators, a 4–8 GHz band-pass filter, and a +44 dB high-electron-mobility-transistor (HEMT) amplifier before returning to the same MW-FEM input channel, where it is digitized and demodulated internally. We use the low-frequency front end module (LF-FEM) to bias the flux lines up to 0.5 V.

\section{Dynamical blockade at microsecond timescales}
\begin{figure*}[h]
\includegraphics[width=1.0\textwidth]{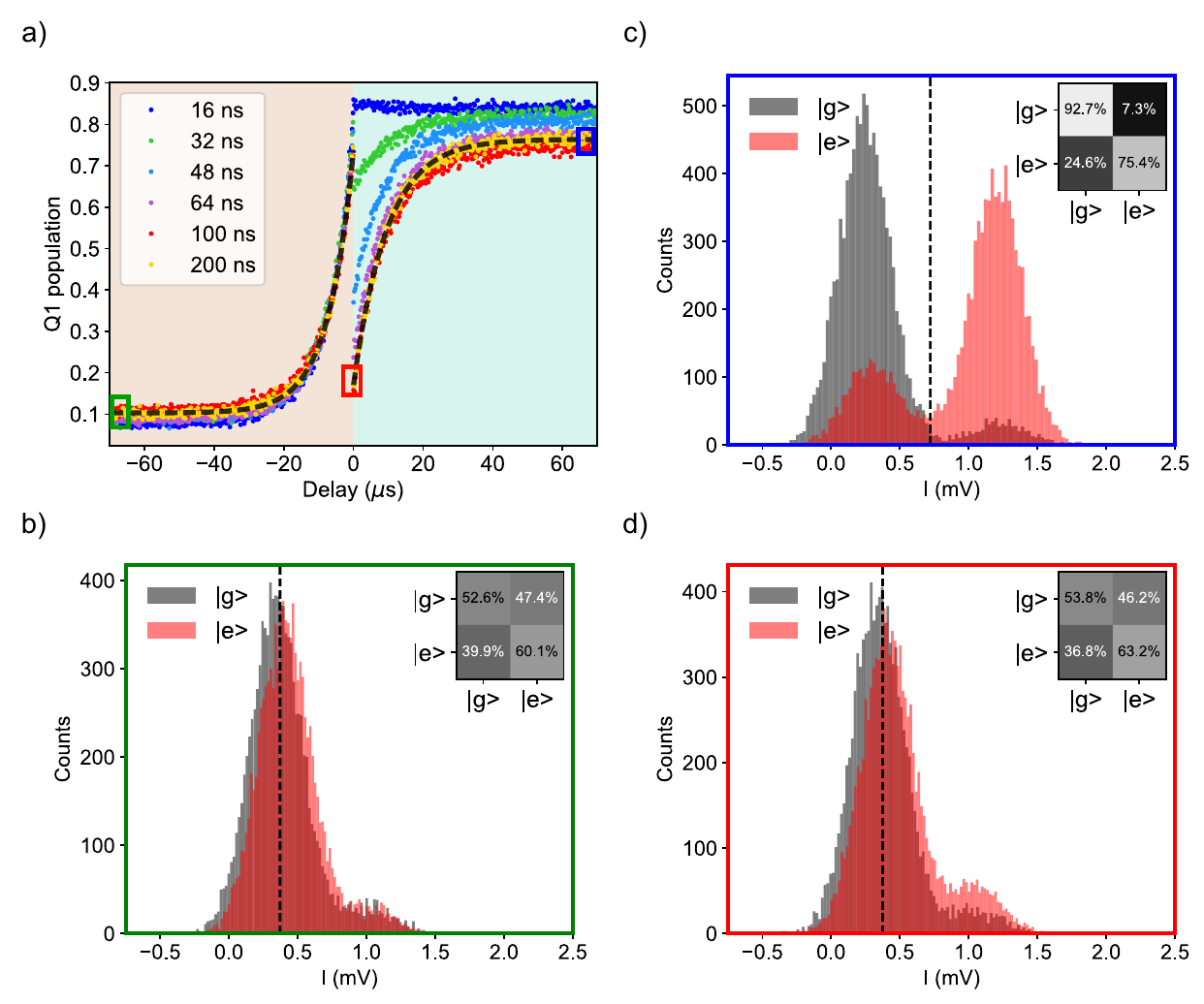}
\caption{\textbf{(a)} Q1 population as a function of the delay between the excitation pulses of different lengths. The black dashed lines mark the fit to the yellow data ($\pi$-pulse length of 200 ns), from which the decay times are extracted. For the same pulse length, the other panels show the readout histograms of the ground (light gray) and excited (light red) state in the rotated IQ plane at selected delays: -75 $\mu$s \textbf{(b)}, 100 ns \textbf{(d)} and 75 $\mu$s \textbf{(c)}. The insets show the readout fidelity matrix.}
\label{fig:blockade_long_SI}
\end{figure*}

To support the results provided in the main text, we show here in Fig.~\ref{fig:blockade_long_SI}a additional data for the dynamical blockade experiment, conducted exactly like that discussed in Fig.~3 of the main text but with longer delay between the excitation pulses.
For negative delays, the excitation pulse sequence corresponds to a standard $T_1$ measurement and, indeed, the population of Q1 decays with a typical timescale of $T = 7.35~\mu$s, consistent with the measured qubit lifetime (see Table \ref{tab:qubit_parameters_main}). For positive delays, we see a degradation of the blockade as Q2 naturally relaxes to the ground state, allowing the population of Q1 to gradually reach the excited state on a timescale of $T = 9.57~\mu$s, consistent with the measured Q2 lifetime (see Table \ref{tab:qubit_parameters_main}).

The readout histograms shown in panels b-c-d represent the state of Q1 at different delays (see also Fig.~\ref{fig:RO_SI} below). For a delay of -75 $\mu$s (panel b, green), most of the excited state counts overlaps with those of the ground state, as expected from a qubit subject to $T_1$ decay. Conversely, for 75 $\mu$s (panel c, blue), the blockade has already been released and the histograms show a typical state separation in the readout plane. Finally, for a short delay of 100 ns (panel d, red), most of the excited state counts are assigned to the ground state because of the effect of the blockade. In each figure, the inset shows the readout fidelity matrix, where each element $ij$ is the probability of preparing Q1 in $\ket{i}$ and measuring it in $\ket{j}$.

\section{Q1 Readout statistics}
\begin{figure*}[h]
\includegraphics[width=0.7\textwidth]{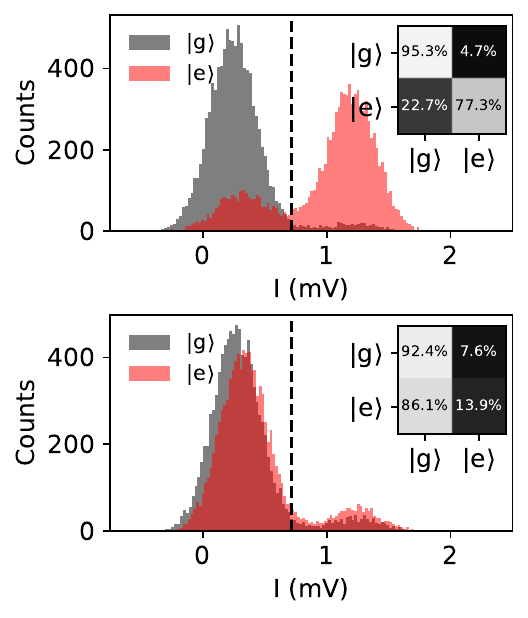}
\caption{The top panel shows the readout state separation of Q1 along the rotated I variable when Q2 is in the ground state. Similarly, the bottom panel shows the same quantities when Q2 is excited 100 ns before Q1. Both qubits are excited with a 100 ns $\pi$ pulse.}
\label{fig:RO_SI}
\end{figure*}

Figure \ref{fig:RO_SI} presents the results of a typical readout operation on Q1, along the rotated I variable, when the qubit is excited with a 100 ns pulse. The top panel shows the readout histograms when Q2 is in the ground state. The $ge$ and $ee$ element of the readout matrix in the inset give the population of Q1 at large negative and positive delay respectively. This behavior is indeed observed in the red data in Fig.~\ref{fig:blockade_long_SI}, where the mismatch between the population at large negative (positive) delay and the fidelity matrix is due to residual excited state population of Q1 (Q2) 75 $\mu$s after its excitation.

The bottom panel shows the same quantities when Q2 is excited 100 ns before Q1, effectively blocking its excitation. As a result, the excited state histogram is shifted to lower voltages and almost overlaps with the ground state. The $eg$ element of the fidelity matrix is now larger than the $ee$ element, showing the effect of the blockade. In particular, the $ee$ element represents the population of Q1 when the delay between the excitation pulses is 100 ns, as is confirmed by looking at Fig.~\ref{fig:blockade_long_SI}.

\section{Extraction of the exchange coupling $J$}
\begin{figure*}[h]
\includegraphics[width=0.8\textwidth]{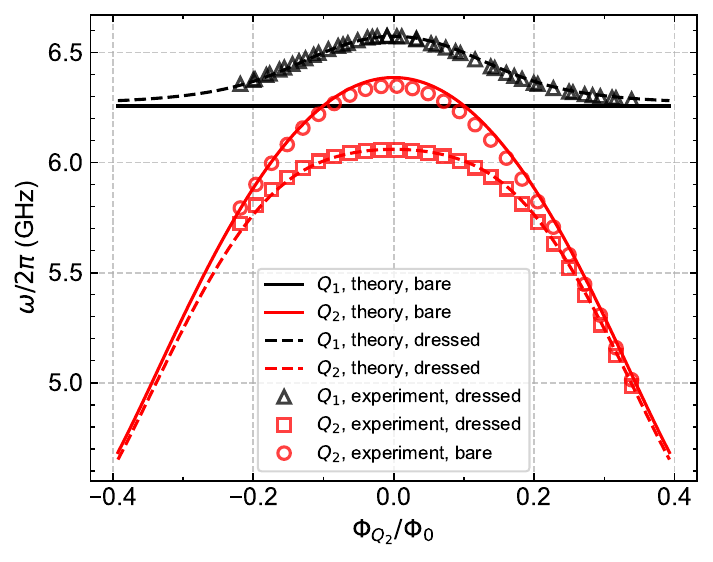}
\caption{The measured excitation spectra for qubits Q1, Q2 vs. flux on Q2. Full lines mark the theoretical bare qubits frequencies (01 transitions, with $H_{\text{int}}\approx0$), where the two qubits frequencies are far apart. The black triangles (red squares) mark the measured Q1 (Q2) frequency vs. $\Phi_{Q2}$ while keeping Q1 at 6.2 GHz (i.e. $\Phi_{Q1} = \Phi_{0} / 2$), clearly showing the level repulsion between Q1 and Q2. The experimental data is very well reproduced by the theoretical calculations explained in Sec. \ref{app:effham}, shown by the black (red) dashed line for Q1 (Q2).}
\label{anticrossing_SI}
\end{figure*}

We have also computed the exchange coupling $J$ from spectroscopy data and compared it against the theoretical value extracted from the circuit Hamiltonian, as shown in Fig.~\ref{anticrossing_SI}. Here, we first perform two-tone spectroscopy on Q2 near its upper sweet-spot at 6.35 GHz when Q1 is placed at 9.2 GHz (red circles). In this case, given the large detuning between the qubits, we are effectively measuring the bare frequency of Q2 as a function of its flux, leading to the expected cosine behavior typical of these systems. We then repeat the same measurement with Q1 fixed at 6.27 GHz (red squares), and observe that the frequency of Q2 shifts to lower energies. This is a signature of level repulsion between the qubits, whose frequency becomes dressed by the $J$ interaction as their detuning is small. For this arrangement, we also perform qubit spectroscopy on Q1 and observe a similar level repulsion (black triangles).
This data was also reproduced by the circuit Hamiltonian (full and dashed lines), whose parameters were independently extracted in previous measurements, showing excellent agreement with the measured values. Additionally, we extract the junction asymmetries by fitting the experimental qubit frequencies as a function of the magnetic flux applied to Q2. The asymmetry between the two Josephson junctions $a$ and $b$ forming a SQUID is quantified through the parameter
\begin{equation}
d = \frac{E_{J,a}-E_{J,b}}{E_{J,a}+E_{J,b}},
\end{equation}
where $E_{J,a}$ and $E_{J,b}$ are the Josephson energy of the junctions. By construction, the asymmetry satisfies $|d|<1$. In the presence of a finite asymmetry, the effective Josephson energy of the SQUID becomes
\begin{equation}
E_J^{\mathrm{eff}}(\Phi)
= E_{J,\Sigma}
\sqrt{
\cos^2\!\left(\frac{\pi \Phi}{\Phi_0}\right)
+ d^2 \sin^2\!\left(\frac{\pi \Phi}{\Phi_0}\right)
},
\end{equation}
where $E_{J,\Sigma}=E_{J,a}+E_{J,b}$. As a consequence, the effective Josephson energy does not vanish at half-integer flux bias, giving rise to a finite minimum qubit frequency, commonly referred to as the lower sweet-spot.

The extracted asymmetries are 0.481 for Q1 and 0.475 for Q2, in excellent agreement with the design values ($\simeq 0.5$).

We can then extract $J$, which is equivalent to the half of the minimum energy splitting between the symmetric and antisymmetric single-excitation eigenstates,
\begin{equation}
E_{\pm} = \frac{E_{01} \pm E_{10}}{\sqrt{2}},
\end{equation}
so that
\begin{equation}
2J = \min \left( E_{+} - E_{-} \right).
\end{equation}
This minimum is found at a $\Phi_{Q_2}$ value of $-0.1 \Phi_{0}$, i.e. at the crossing between the bare frequencies of the two qubits, where $2J \simeq 491~\mathrm{MHz}$.

\section{Additional data from Chip~2}

\subsection*{Experimental Device and Measurement Setup}

\begin{figure}[h!]
    \centering
    \includegraphics[width=1.0\linewidth]{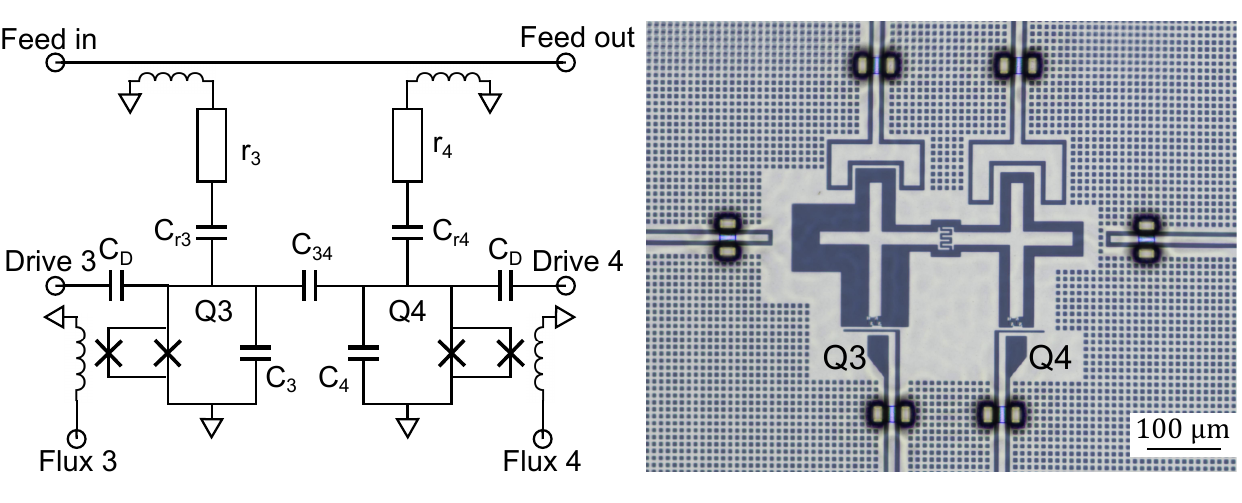}
    \caption{(a) Circuit diagram of the coupled qubits and the two readout resonators. The design value of the shunt capacitances are $C_3=45$ fF and $C_4=55$ fF, while those between the qubits and their resonators are $C_{r3} = C_{r4} = 2.2$ fF. The coupling capacitance between the qubits is $C_{34}=5.5$ fF, while $C_D = 60$ aF. (b) Optical image of the tested device showing the two X-mon qubits with their flux (bottom) and drive (sides) lines and couplings to the readout resonators (top). The capacitive coupling between the qubits is engineered with an interdigitated finger capacitor.}
    \label{fig:device}
\end{figure}

A schematic of the experimental device and an optical micrograph of the fabricated chip are shown in Fig.~\ref{fig:device}.
\begin{figure*}[h]
\includegraphics[width=0.8\textwidth]{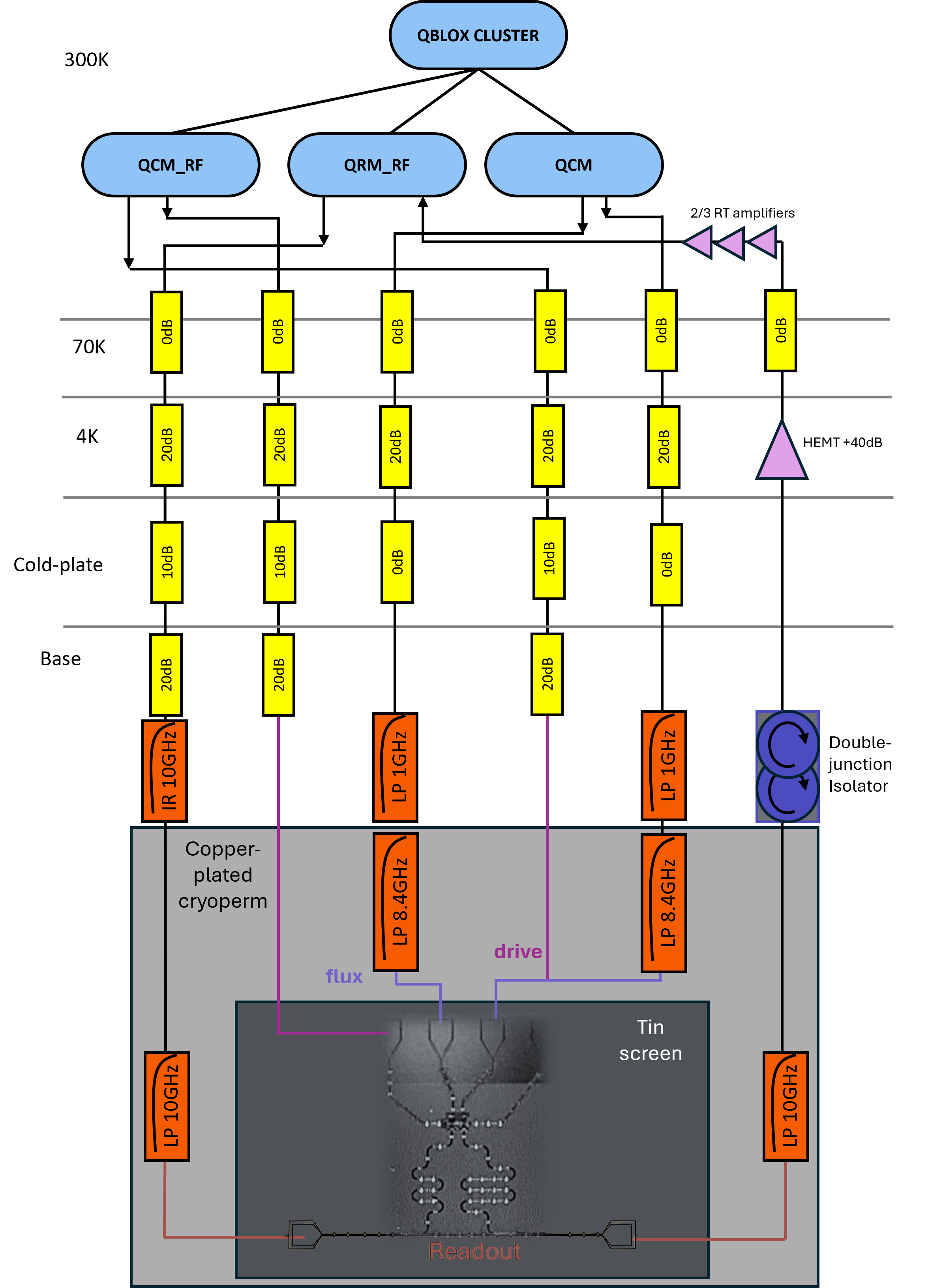}
\caption{\label{fig:Triton} The measurement setup for the second device. The arrows near electronics modules mark the direction of the signal flow.}
\end{figure*}
The device consists of two capacitively coupled, flux-tunable transmon qubits of the X-mon architecture \cite{Barends2013}. Each transmon is implemented as two parallel Al/AlO$_x$/Al Josephson junctions shunted by a capacitor patterned in aluminum on a high-resistivity silicon substrate ($\rho > 10~\text{k}\Omega\cdot\text{cm}$).

The two qubits, denoted Q3 and Q4, are positioned in close proximity to realize a direct capacitive coupling through a lithographically defined finger capacitor, labeled $C_{34}$ in Fig. \ref{fig:device}. The typical feature size of the capacitor fingers is approximately $2~\mu$m, with a total qubit footprint of roughly $200\times250~\mu\text{m}^2$. This coupling allows for the coherent exchange of excitations between the qubits while maintaining independent control of their transition frequencies. 
Each qubit is individually flux-tuned through a coplanar-waveguide (CPW) on-chip shorted near the device  (Flux 3 and Flux 4) that provides a local magnetic flux bias to its respective superconducting quantum interference device (SQUID) loop. 

Each transmon is capacitively coupled to a CPW readout resonator through coupling capacitors $C_{r3}$ and $C_{r4}$, with coupling strengths designed to achieve dispersive readout while minimizing measurement-induced dephasing. The resonators are multiplexed on a common feedline for heterodyne detection. 

The diagram of the setup used for this second device is shown in Fig.~\ref{fig:Triton}. The sample is wire-bonded on a home-made PCB and installed on customized sample holder, anchored at the coldest stage of a Triton400 dilution refrigerator, and screened by multiple stages for both magnetic and radiative protection.

The input microwave line, labeled ``Feed in,'', is connected at room temperature to the output channel of a QRM\_RF module from Qblox, and delivers the measurement tone. The ``Feed out'' port collects the transmitted signal, which is then amplified by a cryogenic high-electron-mobility transistor (HEMT) amplifier at the 4~K stage and further amplified at room temperature before demodulation through the QRM\_RF. Additionally, two independent microwave drive lines (Drive 3 and Drive 4) provide local control over single-qubit rotations via resonant microwave pulses (QCM\_RF modules by Qblox). Finally, flux control is achieved at room temperature by QCM modules. Synchronization and connection with a host PC is guaranteed by a Qblox cluster. Further details on the filtering, attenuation, and screening are reported in Fig.~\ref{fig:Triton}.

At their flux-insensitive sweet-spots, the measured transition frequencies are
$\omega_{3}/(2\pi) = 7.9104~\text{GHz}$ and $\omega_{4}/(2\pi) = 5.5667~\text{GHz}$,
and the corresponding readout resonators are at
$\omega_{r3}/(2\pi) = 6.7070~\text{GHz}$ and $\omega_{r4}/(2\pi) = 4.9688~\text{GHz}$.
The anharmonicities are
$\alpha_{3}/(2\pi) \approx -310~\text{MHz}$ and $\alpha_{4}/(2\pi) \approx -305~\text{MHz}$,
consistent with the transmon regime~\cite{Koch2007}. Energy-relaxation times are typically 
$T_{1} = 15$--$20~\mu\text{s}$,
and spin-echo coherence times are
$T_{2}^{\text{echo}} = 18$--$25~\mu\text{s}$.
A complete list of the relevant device parameters---including capacitances, frequencies, anharmonicities, and coherence times---is summarized in Table \ref{tab:qubit_parameters_SI}.

\begin{table}[h]
\caption{\label{tab:qubit_parameters_SI}%
System parameters.}
\begin{ruledtabular}
\begin{tabular}{rcc}
 & \textrm{Q3}& \textrm{Q4}\\
\colrule
$\omega/(2\pi)$ (GHz) & 7.9104 & 5.5667 \\
$\alpha/(2\pi)$ (MHz) & -324 & -288 \\
$\omega_r/(2\pi)$ (GHz)& 6.7070 & 4.9688 \\
g$/(2\pi)$ (MHz) & 40.50 & 29.29 \\
$\chi/(2\pi)$ (kHz) & 292 & 707 \\
T$_1$ ($\mu$s) & 2.68 & 11.92 \\
T$_2^*$ ($\mu$s) & 1.98 & 3.53 \\
T$_2$ ($\mu$s) & 2 & 3.11 \\
\end{tabular}
\end{ruledtabular}
\end{table}

\subsection{ZZ measurement protocol via Ramsey and Echo}
\begin{figure*}
\includegraphics[width=1.0\textwidth]{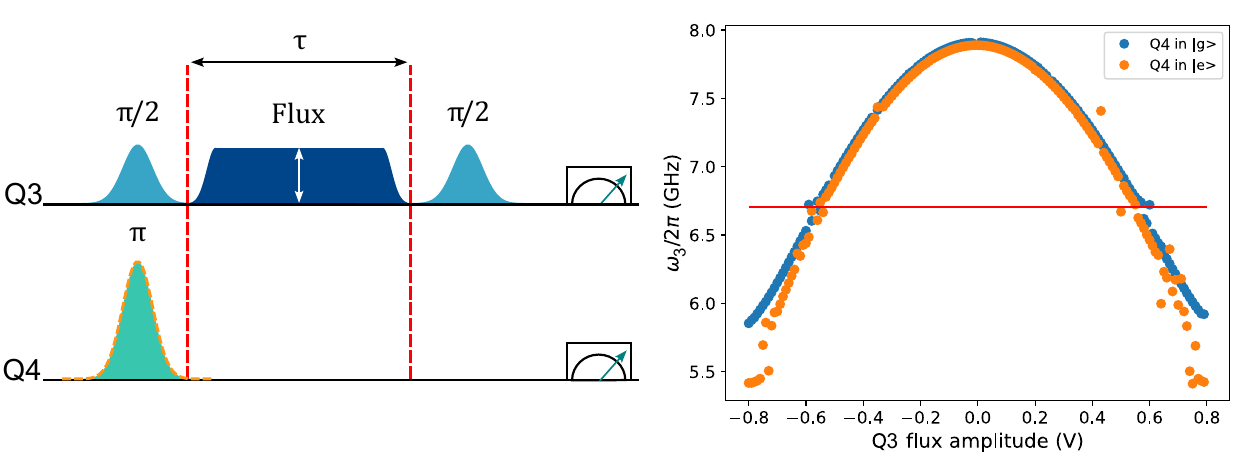}
\caption{\label{fig:Ramsey} Calibration of the amplitude of the flux pulse on Q3. The red horizontal line shows the spectral position of the readout resonator of Q3. The ZZ is then computed as the difference between the blue data points (acquired with Q4 in the ground state) and the orange ones (with Q4 in the excited state).}
\end{figure*}

The measurement protocol to characterize the ZZ interaction is shown in Fig. \ref{fig:Ramsey} and consists in measuring the frequency of Q3 with a Ramsey sequence both when Q4 is in the ground state and in the excited state, similarly to what was performed on the other device. We repeat this measurement at different frequencies of Q3 by applying a flux pulse during the Ramsey sequence, while Q4 is kept at its upper sweet-spot frequency of 4.97 GHz (see Fig.~\ref{fig:Ramsey}). By changing the amplitude of this flux pulse we are able to characterize Q3 at different frequencies, going from its upper sweet-spot at 7.91 GHz all the way down to below 6 GHz.
The results of these measurements are plotted in Fig.~\ref{fig:Ramsey}, where the blue points show the frequency of Q3 measured when Q4 is in the ground state and the orange ones the frequency of Q3 when Q4 is in the excited state. Indeed, the orange points are consistently below the blue ones, confirming the theoretical prediction that the ZZ interaction shifts the resonant frequency of Q3 to lower energies. Moreover, this shift is larger as the detuning between the qubits is reduced, i.e. at larger flux amplitudes, and $\zeta$ becomes stronger (see Equation 5 in the main text).

\begin{figure*}
\centering
\includegraphics[width=1.0\textwidth]{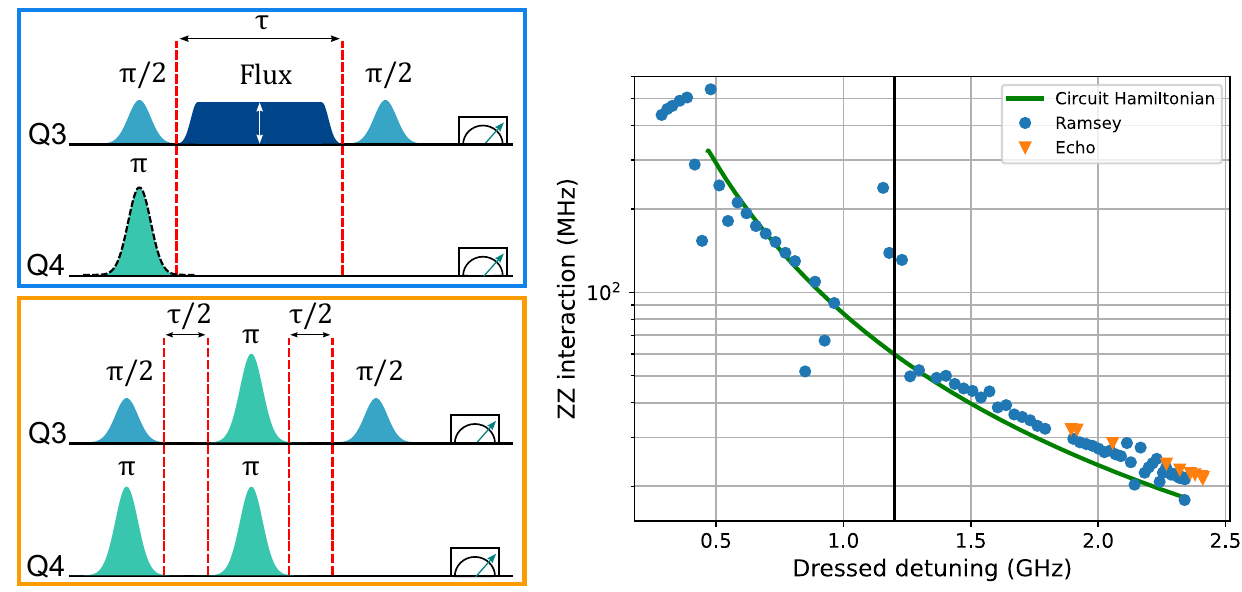}
\caption{\label{fig:ZZvsDelta_Naples} Measured ZZ interaction as a function of the frequency detuning between the qubits. The blue data points show the ZZ measured with a Ramsey experiment (blue pulse sequence on the top left), while the orange ones are extracted from the modified Echo sequence shown on the bottom left \cite{Rol2020}. Additionally, we show in green the ZZ extracted from the circuit Hamiltonian. The black line labels the spectral position of the readout resonator of Q3, which affects the measurements during the Ramsey protocol.}
\end{figure*}

Finally, Fig.~\ref{fig:ZZvsDelta_Naples} shows, in blue, the extracted ZZ interaction as a function of the dressed detuning between Q3 and Q4. Additionally, we show in orange the ZZ interaction measured via a modified Echo sequence \cite{Rol2020}, which is in good agreement with the value measured via Ramsey. We have found, however, that this method is less resilient for detuning between qubits smaller than 2 GHz. We note here that we could measure the ZZ interaction via the Ramsey method to higher frequency values than in the main text, as in this setup the time resolution of the electronics reached 1 ns. 

\subsection{Blockade calibration}
\begin{figure}
\includegraphics[width=1.0\textwidth]{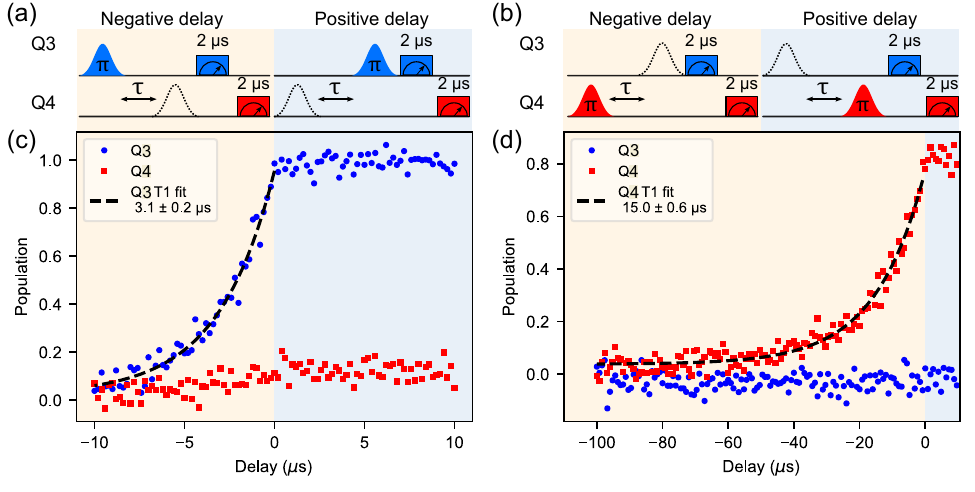}
\caption{\label{fig:blockade_calib} Blockade experiment calibration. Left (right):  $\pi$-pulse on Q3 (Q4), and no pulse on Q4 (Q3). The black dashed lines show the fits to the decaying populations, from which the relaxation times of the qubits are extracted. The pulse sequence used for the measurements is shown on top (a, b), with the dashed pulse indicating the missing drive signal. On the left-hand side (c), {the population of Q4 increases to about 0.1 as we cross the delay from negative to positive most likely due to drive crosstalk from Q3}. On the right-hand side (d), Q4 population never reaches 1 as this qubit is always read-out after Q3 and therefore it decays during this first readout operation.} 
\end{figure}

Before performing the blockade experiment, we execute a simpler blockade sequence where either the drive pulse on Q3 or Q4 is missing. The resulting pulse sequences are shown in Fig.~\ref{fig:blockade_calib} together with the measured populations of the qubits. Looking at the blue-shaded area in the left (right) plot,  one sees that the population of Q3 (Q4) is always constant at unity as the qubit is measured immediately after the $\pi$-pulse, while the population of the other qubit is always in $\ket{0}$ as it is not excited. For negative delays (yellow-shaded area), we simply measure the excited state population decay of Q3 in the left plot and Q4 in the right one. A fit to the data yields the $T_1$ values of the qubits, which agree with those measured with the standard protocol and reported in Table \ref{tab:qubit_parameters_SI}.

\subsection{Blockade experiment}
\begin{figure*}
\includegraphics[width=1.0\textwidth]{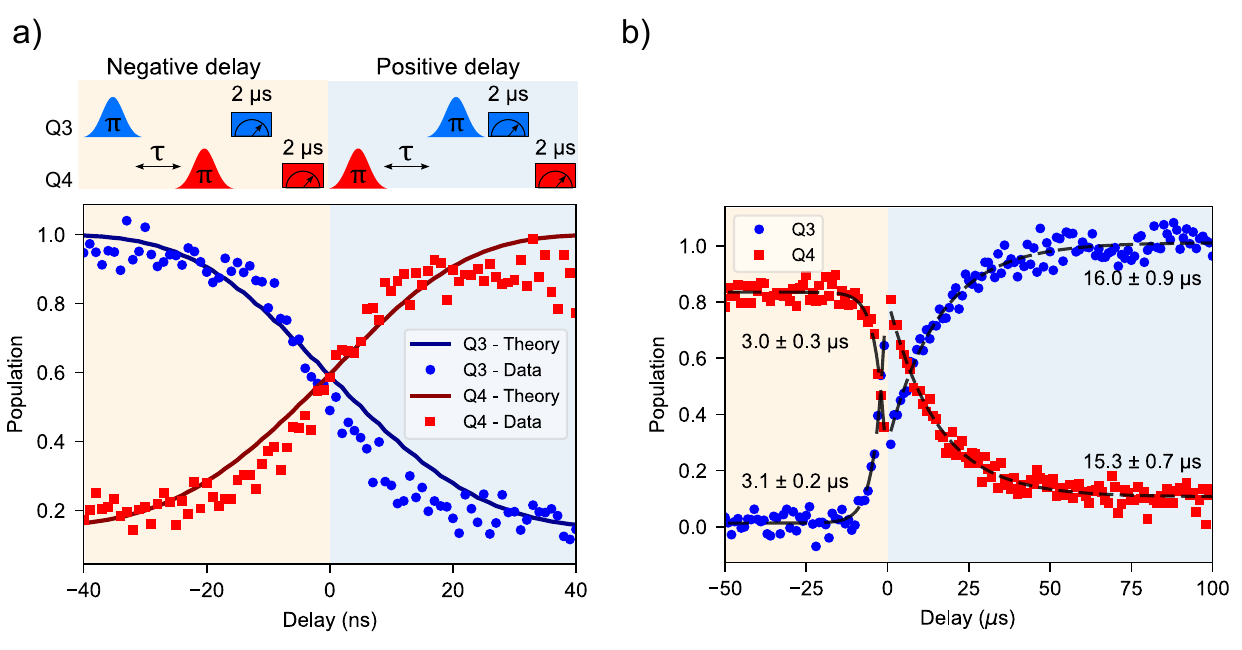}
\caption{
(\textbf{a}) Populations of Q3 (blue) and Q4 (red) versus relative delay in the short-delay regime. Exciting Q4 first (positive delay) suppresses excitation of Q3, demonstrating blockade due to the strong cross-Kerr coupling ($\zeta/(2\pi) = 20~\text{MHz}$). The top inset shows the pulse sequences for Q3 and Q4 illustrating positive and negative excitation delays. Each $\pi$-pulse has a Gaussian envelope ($\sigma = 60~\text{ns}$) and is followed by sequential dispersive readout.
(\textbf{b}) Population dynamics over longer delays show exponential relaxation consistent with the measured qubit lifetimes ($T_{1}^{(3)} = 3.1~\mu$s, $T_{1}^{(4)} = 15.3~\mu$s). 
The observed suppression of excitation confirms the emergence of a Rydberg-type blockade mediated by engineered ZZ interactions.}
\label{fig:blockade_Naples}
\end{figure*}

For a fixed pulse length of 60 ns, we show in Fig.~\ref{fig:blockade_Naples} the population dynamics of Q3 and Q4 as a function of the delay between their excitations, for both short (a) and long (b) delays.
The two populations are clearly anti-correlated, i.e. one falls when the other one rises, showing the effect of the blockade. As for the device discussed in the main text, in panel (a) we reproduce the dynamics through blockade simulations, while in panel (b) we fit the decay of the populations and recover the relaxation times of the qubits.

\subsection{Readout statistics}
\begin{figure}
\includegraphics[width=1.0\textwidth]{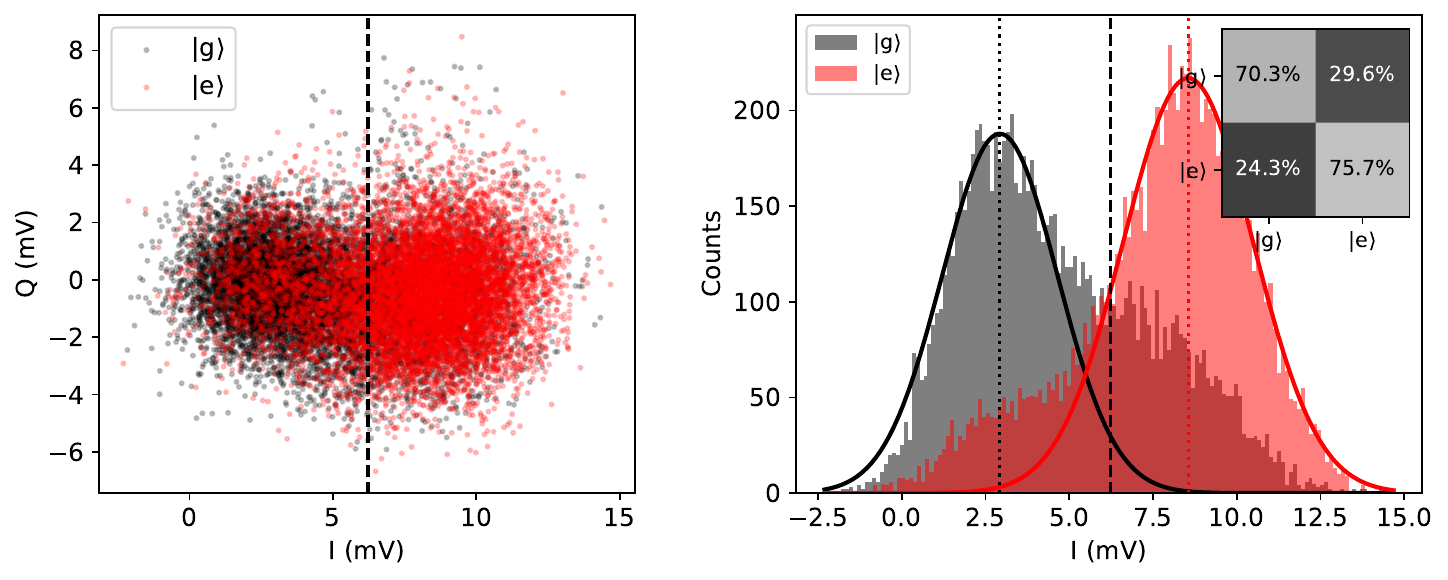}
\caption{\label{fig:SSRO_Q3} Single shot read out statistics for Q3. The ground-state data was fitted in the range $I<$4.5 mV and the excited-state data in the range $I>$6 mV. These ranges were chosen to highlight deviations from the expected Gaussian distributions. The long tail in the ground-state data indicates that the qubit was excited during readout, while the short tail in the excited-state data suggests that the qubit decayed during readout.}
\end{figure}

\begin{figure}
\includegraphics[width=1.0\textwidth]{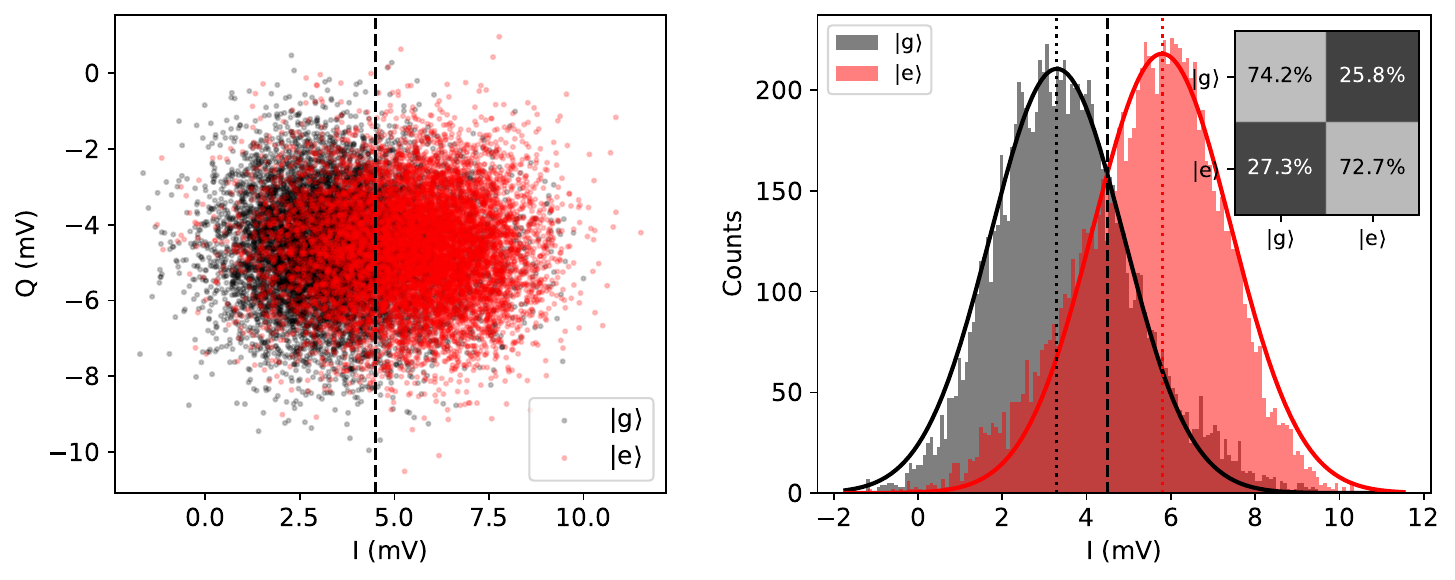}
\caption{\label{fig:SSRO_Q4} Single shot read out statistics for Q4.}
\end{figure}

We show in Figs.~\ref{fig:SSRO_Q3} and \ref{fig:SSRO_Q4} the readout blobs and histograms of Q3 and Q4, respectively, when their partner was in the ground state. In both figures, the inset shows the readout fidelity matrix.

\newpage

\section{Black-box quantization via Foster synthesis}
\subsection{Linearization and port admittance}

Black-box quantization starts from the linear response of the electromagnetic environment seen by each Josephson junction (JJ) port \cite{Nigg2012,Solgun2014}. 
Each JJ is linearized by replacing the cosine element with an inductance
\begin{equation}
L_{J,j}=\frac{\Phi_0^2}{E_{J,j}},\qquad \Phi_0=\frac{\hbar}{2e},
\end{equation}
(optionally including its junction capacitance if relevant). The resulting linear circuit is fully characterized by the driving-point admittance at the JJ port,
\begin{equation}
Y_j(\omega)=\frac{I_j(\omega)}{V_j(\omega)},
\end{equation}
obtained from circuit analysis or EM simulation \cite{Nigg2012,Solgun2014,Solgun2015}.

\subsection{Rational fit and Foster decomposition}

To obtain a finite modal representation, $Y_j(\omega)$ (or $Z_j(\omega)=1/Y_j(\omega)$) is fitted by a passive (positive-real) rational function in $s=i\omega$ \cite{Solgun2014,Solgun2015}. 
In practice, this is commonly done with vector fitting \cite{Gustavsen1999}, yielding poles and residues that define the normal modes.

In the weakly dissipative limit Foster synthesis represents the impedance as a series of parallel resonators (Foster I form) \cite{Foster1924,Nigg2012}:
\begin{equation}
Z_j(\omega)\simeq \sum_m Z_m(\omega), 
\qquad 
Z_m(\omega)=\left(\frac{1}{i\omega L_m}+i\omega C_m+\frac{1}{R_m}\right)^{-1}.
\end{equation}
Each mode has $\omega_m=1/\sqrt{L_m C_m}$ and (if losses are kept) $\kappa_m=1/(R_m C_m)$.
Equivalently, one can extract $(\omega_m,\kappa_m)$ from the complex poles $\zeta_m=\omega_m+i\kappa_m/2$ and obtain effective modal parameters from residues/derivatives of the fitted admittance \cite{Nigg2012,Solgun2014}. 
Software packages such as \texttt{QuCAT} automate this step~\cite{GelySteele2020QuCAT}.

\subsection{Quantization and restoration of Josephson nonlinearity}

Neglecting dissipation ($R_m\to\infty$), the Foster modes are quantized as harmonic oscillators \cite{Nigg2012}:
\begin{equation}
\hat{H}_{\mathrm{lin}}=\sum_m \left(\frac{\hat{\Phi}_m^2}{2L_m}+\frac{\hat{Q}_m^2}{2C_m}\right)
=\sum_m \hbar\omega_m\left(\hat{a}_m^\dagger\hat{a}_m+\tfrac12\right),
\end{equation}
with $[\hat{\Phi}_m,\hat{Q}_n]=i\hbar\delta_{mn}$ and
\begin{equation}
\hat{\Phi}_m=\Phi_{\mathrm{zpf},m}(\hat{a}_m+\hat{a}_m^\dagger),
\qquad
\Phi_{\mathrm{zpf},m}=\sqrt{\frac{\hbar Z_m}{2}},
\qquad
Z_m=\sqrt{\frac{L_m}{C_m}}.
\end{equation}

The JJ nonlinearity is restored by replacing the linear inductor with the cosine potential while subtracting the quadratic part already included in $\hat{H}_{\mathrm{lin}}$ \cite{Nigg2012,Solgun2014}:
\begin{equation}
\hat{H}=\sum_m \hbar\omega_m \hat{a}_m^\dagger\hat{a}_m
+\sum_j E_{J,j}\!\left[1-\cos\hat{\varphi}_j-\frac{\hat{\varphi}_j^{\,2}}{2}\right],
\end{equation}
where the phase across junction $j$ is expanded in the normal-mode basis,
\begin{equation}
\hat{\varphi}_j=\sum_m \varphi_{\mathrm{zpf},m,j}\,(\hat{a}_m+\hat{a}_m^\dagger).
\end{equation}
In the weakly anharmonic regime ($\varphi_{\mathrm{zpf},m,j}\ll 1$), keeping the leading (quartic) term and applying the rotating-wave approximation gives the Kerr Hamiltonian \cite{Nigg2012}:
\begin{equation}
\hat{H}\approx \sum_m \hbar\omega_m \hat{a}_m^\dagger\hat{a}_m
-\frac12\sum_m |\alpha_m|\,\hat{a}_m^{\dagger 2}\hat{a}_m^{2}
-\sum_{m<n}\chi_{mn}\, \hat{a}_m^\dagger\hat{a}_m\,\hat{a}_n^\dagger\hat{a}_n,
\label{eq:H_kerr}
\end{equation}
with
\begin{equation}
|\alpha_m|=\sum_j \frac{E_{J,j}}{2}\,\varphi_{\mathrm{zpf},m,j}^4,
\qquad
\chi_{mn}=\sum_j E_{J,j}\,\varphi_{\mathrm{zpf},m,j}^2\,\varphi_{\mathrm{zpf},n,j}^2
=2\sum_j \sqrt{\alpha_{m,j}\alpha_{n,j}},
\end{equation}
(where $\alpha_{m,j}=\frac{E_{J,j}}{2}\varphi_{\mathrm{zpf},m,j}^4$).
This provides a compact route from a fitted linear admittance to an effective multimode Hamiltonian.

\subsubsection*{Effective Hamiltonian - analytical formula}
\label{app:effham}
We now want to show in which sense Eq.~\eqref{eq:H_kerr} can be interpreted, when projected to the qubit state, as the Hamiltonian 
\begin{equation}
    \label{eq:staticHsi}
    \hat{H}_0 := \beta_0 {\bm 1}^{(1)} \otimes {\bm 1}^{(2)} +  \beta_1 {\bm 1}^{(1)} \otimes \hat{Z}_2 +  \beta_2 \sigma_x^{(1)} \otimes \hat{X}_2 +  \beta_3 \sigma_y^{(1)} \otimes \sigma_y^{(2)} +  \beta_4 \sigma_z^{(1)} \otimes {\bm 1}^{(2)} +  \beta_5 \sigma_z^{(1)} \otimes \sigma_z^{(2)},
\end{equation}
with $\beta_2=\beta_3$.
Note that the Hamiltonian of Eq.~\eqref{eq:H_kerr} contains more terms than those of Eq.~\eqref{eq:staticHsi}. However, as we show below, the terms of the form $\sigma_{x}\sigma_{x}$ and $\sigma_{y}\sigma_{y}$ are uninfluential (e.g.  see the later discussion in Sec \ref{app:effblockade}), while terms of the form $\sigma_x\sigma_y$ are sufficiently small to be negligible. The spectrum shown in Fig.~\ref{anticrossing_SI} is also obtained by diagonalizing such Hamiltonian.

We begin with the multimode Kerr Hamiltonian of Eq.~\eqref{eq:H_kerr}:
\begin{equation}
\hat{H} =
\sum_{m} \hbar \omega_m \hat{a}_m^{\dagger} \hat{a}_m
-\frac{1}{2}\sum_{m} |\alpha_m| \hat{a}_m^{\dagger 2}\hat{a}_m^{2}
-\sum_{m<n} \chi_{mn} \hat{a}_m^{\dagger}\hat{a}_m
\hat{a}_n^{\dagger}\hat{a}_n,
\label{eq:H_kerr_supp}
\end{equation}
and include an exchange-type coupling between modes,
\begin{equation}
\hat{H}_{\mathrm{int}} = g \left( \hat{a}_1^{\dagger}\hat{a}_2 + \hat{a}_1\hat{a}_2^{\dagger} \right).
\end{equation}
We assume the mode is weakly anharmonic (implying that we can approximate it as a transmon) with anharmonicity $\alpha_m$, frequency $\omega_m$, and detuning
$\Delta = \omega_1 - \omega_2$.

To capture the dispersive cross-coupling, we truncate each oscillator to the three lowest levels
$\{|0\rangle, |1\rangle, |2\rangle\}$,
with matrix elements
$a|1\rangle = |0\rangle$ and $a|2\rangle = \sqrt{2}\,|1\rangle$. This is the minimal truncation that exhibits ZZ coupling \cite{DiCarlo2009}.
The unperturbed (bare) energies are
\begin{align}
E_{00} &= 0, &
E_{10} &= \omega_1, &
E_{01} &= \omega_2, \notag\\
E_{11} &= \omega_1 + \omega_2, &
E_{20} &= 2\omega_1 - |\alpha_1|, &
E_{02} &= 2\omega_2 - |\alpha_2|.
\end{align}
The relevant interaction matrix elements in this truncated space are
\begin{align}
\langle 01|\hat{H}_{\mathrm{int}}|10\rangle &= g, &
\langle 20|\hat{H}_{\mathrm{int}}|11\rangle &= \sqrt{2}\,g, &
\langle 02|\hat{H}_{\mathrm{int}}|11\rangle &= \sqrt{2}\,g.
\end{align}

We perform a Schrieffer--Wolff (SW)\cite{DevoretSchoelkopf2013} transformation to eliminate the non-computational states
$|20\rangle$ and $|02\rangle$ perturbatively to second order in $g$.
The relevant detunings are
\begin{align}
E_{11} - E_{20} &= -(\Delta - |\alpha_1|), &
E_{11} - E_{02} &=  (\Delta + |\alpha_2|).
\end{align}
The second-order correction to the energy of $|11\rangle$ is
\begin{equation}
\delta E_{11}^{(2)} = 
\frac{2g^2}{E_{11}-E_{20}} + \frac{2g^2}{E_{11}-E_{02}}
= -\frac{2g^2}{\Delta - |\alpha_1|} + \frac{2g^2}{\Delta + |\alpha_2|}.
\end{equation}
The single-excitation states acquire dispersive shifts which are given by the quantities
\begin{equation}
\delta E_{10}^{(2)} = \frac{g^2}{\Delta}, \qquad
\delta E_{01}^{(2)} = -\frac{g^2}{\Delta}.
\end{equation}

The effective cross-Kerr or ZZ rate is defined as the differential energy shift
\begin{equation}
\zeta = 
\left(E_{11} - E_{10} - E_{01} + E_{00}\right)_{\mathrm{dressed}} =
2g^2\!\left(
\frac{1}{\Delta + |\alpha_2|}
- \frac{1}{\Delta - |\alpha_1|}
\right),
\label{eq:zeta}
\end{equation}
recovering Eq.\ 5 in the main text. If a bare cross-Kerr term $-\chi\,\hat{n}_1\hat{n}_2$ is included in Eq.~\eqref{eq:H_kerr_supp},
it contributes an additional $\zeta_{\mathrm{bare}} = -\chi$, such that
\begin{equation}
\zeta = -\chi +
2g^2\!\left(
\frac{1}{\Delta + |\alpha_2|}
- \frac{1}{\Delta - |\alpha_1|}
\right).
\label{eq:zeta_total}
\end{equation}

It is interesting to see that at high detuning the first correction to $\zeta$ in $1/\Delta$ is independent from the anharmonicity of the qubits $A_i$, e.g.:
\begin{align}
\zeta_{\Delta \gg |\alpha_i|}
&\simeq-\chi+
2 g^{2}(|\alpha_{1}|+|\alpha_{2}|)\left[
\frac{1}{\Delta^{2}}
+ \frac{|\alpha_{2}|-|\alpha_{1}|}{\Delta^{3}}
+ \frac{|\alpha_{2}|^{2}-|\alpha_{1}||\alpha_{2}|+|\alpha_{1}|^{2}}{\Delta^{4}}
+ O\!\left(\frac{1}{\Delta^{5}}\right)
\right].
\end{align}

In the special case $\alpha_1=\alpha_2=\alpha$, we obtain:
 $$\zeta_{\Delta \gg |\alpha_i|} \simeq -\chi + 4g^2|\alpha|\left[\frac{1}{\Delta^2} + \frac{\alpha^2}{\Delta^4} + \mathcal{O}(\frac{1}{\Delta^{5}})\right].$$

We have also compared the $\zeta$ value predicted by Eq.\ \ref{eq:zeta} with the measured data, and found a discrepancy in the order of 10\% also at large $\Delta$. This mismatch likely stems from the elimination of higher order terms that are often dropped out with rotating-wave approximation~\cite{Solgun_PRApp_2022}.

Restricting to the computational basis 
$\{|00\rangle, |01\rangle, |10\rangle, |11\rangle\}$,
the effective static Hamiltonian takes the form
\begin{equation}
\hat{H}_{\mathrm{eff}} =
\frac{\tilde{\omega}_1}{2}\sigma_z^{(1)}
+ \frac{\tilde{\omega}_2}{2}\sigma_z^{(2)}
+ \frac{J}{2}\left(\sigma_x^{(1)}\sigma_x^{(2)} + \sigma_y^{(1)}\sigma_y^{(2)}\right)
+ \frac{\zeta}{4}\,\sigma_z^{(1)}\sigma_z^{(2)}
+ \text{const.}
\label{eq:H_eff_supp}
\end{equation}
Here, $J \approx g$ is the dressed exchange rate, and the ZZ coefficient is given by Eq.~\eqref{eq:zeta_total}.
In the notation of Eq.~(2) of the main text,
\begin{align}
\beta_2 &= \beta_3 = \tfrac{J}{2}, \qquad
\beta_5 = \tfrac{\zeta}{4}, \\
\beta_1 &= \tfrac{1}{2}\tilde{\omega}_2, \qquad
\beta_4 = \tfrac{1}{2}\tilde{\omega}_1, \qquad
\beta_0 = \text{constant.}
\end{align}

Thus, truncating the multimode Kerr Hamiltonian to third order (qutrit level) and projecting onto the qubit subspace naturally yields the effective static Hamiltonian
\begin{equation}
\hat{H}_0 =
\beta_0\, {\bm 1}^{(1)}\!\otimes\!{\bm 1}^{(2)}
+ \beta_1\, {\bm 1}^{(1)}\!\otimes\!\sigma_z^{(2)}
+ \beta_2\, \sigma_x^{(1)}\!\otimes\!\sigma_x^{(2)}
+ \beta_3\, \sigma_y^{(1)}\!\otimes\!\sigma_y^{(2)}
+ \beta_4\, \sigma_z^{(1)}\!\otimes\!{\bm 1}^{(2)}
+ \beta_5\, \sigma_z^{(1)}\!\otimes\!\sigma_z^{(2)}.
\label{eq:heff_simp}
\end{equation}
This derivation explicitly shows how the virtual $|2\rangle$ levels of each oscillator generate an effective ZZ coupling even in the absence of an explicit cross-Kerr term.

We now show that such interaction shifts the energies in the computational basis of the qubit state. We consider the computational basis $\{\ket{00},\ket{01},\ket{10},\ket{11}\}$ with
$\sigma_z\ket{0}=+\,\ket{0}$ and $\sigma_z\ket{1}=-\,\ket{1}$.
Note that $\braket{i|\sigma_x|i}=\braket{i|\sigma_y|i}=0$ for $i\in\{0,1\}$, so
the XX and YY terms do not contribute to the diagonal matrix elements in this basis.
Therefore, for $\ket{ij}$ we have
\begin{equation}
\begin{aligned}
\langle ij|\hat H_0|ij\rangle
&= \beta_0
+ \beta_1\,\langle j|\sigma_z|j\rangle
+ \beta_4\,\langle i|\sigma_z|i\rangle
+ \beta_5\,\langle i|\sigma_z|i\rangle\langle j|\sigma_z|j\rangle\,.
\end{aligned}
\end{equation}
Evaluating for each basis state gives
\begin{equation}
\begin{aligned}
E_{\ket{00}} &= \beta_0 + \beta_1 + \beta_4 + \beta_5,\\
E_{\ket{01}} &= \beta_0 - \beta_1 + \beta_4 - \beta_5,\\
E_{\ket{10}} &= \beta_0 + \beta_1 - \beta_4 - \beta_5,\\
E_{\ket{11}} &= \beta_0 - \beta_1 - \beta_4 + \beta_5.
\end{aligned}
\end{equation}
After substitution, we obtain
\begin{equation}
\begin{aligned}
\zeta
&= E_{\ket{11}} - E_{\ket{10}} - E_{\ket{01}} + E_{\ket{00}}\\
&= (\beta_0 - \beta_1 - \beta_4 + \beta_5)
  - (\beta_0 + \beta_1 - \beta_4 - \beta_5)
  - (\beta_0 - \beta_1 + \beta_4 - \beta_5)
  + (\beta_0 + \beta_1 + \beta_4 + \beta_5)\\[2pt]
&= 4\,\beta_5.
\end{aligned}
\end{equation}
Thus
\begin{equation}
\zeta = E_{\ket{11}} - E_{\ket{10}}-E_{\ket{01}}  +E_{\ket{00}}\,,
    \label{ZZ strength aiudi0}
\end{equation}
In this parametrization, the ZZ strength is directly proportional to the coefficient
of $\sigma_z^{(1)}\!\otimes\!\sigma_z^{(2)}$. The identity and single-qubit $Z$ terms,
as well as the off-diagonal XX/YY couplings cancel in the combination that defines $\zeta$.

To see why this result is important, let us consider the following observation. We define the conditional single-qubit transition frequencies
\begin{equation}
\omega_1^{(j)} \equiv \frac{E_{\ket{1j}}-E_{\ket{0j}}}{\hbar},\qquad
\omega_2^{(i)} \equiv \frac{E_{\ket{i1}}-E_{\ket{i0}}}{\hbar},\qquad i,j\in\{0,1\}.
\end{equation}
Using the energies from $\hat H_0$,

\begin{align}
E_{\ket{00}} &= \beta_0+\beta_1+\beta_4+\beta_5, \quad
E_{\ket{01}} = \beta_0-\beta_1+\beta_4-\beta_5,\\
E_{\ket{10}} &= \beta_0+\beta_1-\beta_4-\beta_5, \quad
E_{\ket{11}} = \beta_0-\beta_1-\beta_4+\beta_5,
\end{align}

one finds
\begin{equation}
\begin{aligned}
\omega_1^{(0)} &= \frac{E_{\ket{10}}-E_{\ket{00}}}{\hbar}
= -\,\frac{2}{\hbar}\,(\beta_4+\beta_5),\\[2pt]
\omega_1^{(1)} &= \frac{E_{\ket{11}}-E_{\ket{01}}}{\hbar}
= -\,\frac{2}{\hbar}\,(\beta_4-\beta_5),\\[6pt]
\omega_2^{(0)} &= \frac{E_{\ket{01}}-E_{\ket{00}}}{\hbar}
= -\,\frac{2}{\hbar}\,(\beta_1+\beta_5),\\[2pt]
\omega_2^{(1)} &= \frac{E_{\ket{11}}-E_{\ket{10}}}{\hbar}
= -\,\frac{2}{\hbar}\,(\beta_1-\beta_5).
\end{aligned}
\end{equation}
Hence, the \emph{conditional} shifts of each qubit’s frequency (with the other qubit in $\ket{1}$ vs.\ $\ket{0}$) are
\begin{equation}
\Delta\omega_1 \equiv \omega_1^{(1)}-\omega_1^{(0)}=\frac{4\beta_5}{\hbar},\qquad
\Delta\omega_2 \equiv \omega_2^{(1)}-\omega_2^{(0)}=\frac{4\beta_5}{\hbar}.
\end{equation}
Using Eq.~\eqref{ZZ strength aiudi0} (i.e.\ $\zeta=4\beta_5$) this gives the experimental identities
\begin{equation}
\ \zeta=\hbar\,\Delta\omega_1=\hbar\,\Delta\omega_2.
\end{equation}
Thus, under weak-drive spectroscopy, and under the assumption that the Hamiltonian is of the form of Eq.~\eqref{eq:heff_simp}, each qubit’s single-photon line splits into a doublet whose separation is $\zeta/\hbar$; measuring that splitting directly yields the ZZ strength. In Ramsey/echo, preparing one qubit in $\ket{1}$ vs.\ $\ket{0}$ makes the partner acquire an extra phase rate $\Delta\omega_{1,2}=\zeta/\hbar$, providing a time-domain route to the same ZZ calibration (see the following section).

\subsubsection*{Ramsey echo}
In the presence of a static ZZ interaction, the Hamiltonian contains a term
\begin{equation}
\hat{H}_{ZZ} = \hbar\,\frac{\zeta}{4}\,\sigma_z^{(1)}\!\otimes\!\sigma_z^{(2)}.
\end{equation}
For Q1, this implies that its transition frequency depends on the
instantaneous $z$-state of Q2:
\begin{equation}
\omega_1^{(0)} = \omega_1 - \frac{\zeta}{2\hbar},\qquad
\omega_1^{(1)} = \omega_1 + \frac{\zeta}{2\hbar}.
\end{equation}
An analogous relation holds for Q2.
Hence, if Q2 is in $\ket{1}$ instead of $\ket{0}$, Q1's
precession frequency during free evolution changes by
\begin{equation}
\Delta\omega_1 = \omega_1^{(1)} - \omega_1^{(0)} = \frac{\zeta}{\hbar}.
\end{equation}

In a Ramsey experiment, Q1 is prepared in a superposition
$(\ket{0}+\ket{1})/\sqrt{2}$, allowed to evolve freely for a time $\tau$,
and then measured in the equatorial plane.
If Q2 remains in a definite state $\ket{j}$,
the phase accumulated by Bloch vector of Q1 is
\begin{equation}
\phi_1^{(j)} = \omega_1^{(j)}\,\tau.
\end{equation}
The relative phase between the two conditional evolutions is therefore
\begin{equation}
\Delta\phi_1 = (\omega_1^{(1)}-\omega_1^{(0)})\,\tau
= \frac{\zeta\,\tau}{\hbar}.
\end{equation}
Experimentally, this appears as a shift in the Ramsey fringe frequency
depending on the state of the spectator qubit.  Equivalently, if Q2 is
prepared in a superposition, the Ramsey signal of Q1 shows a beating pattern
at frequency $\zeta/\hbar$ corresponding to the accumulation of the entangling phase.

The same effect can be observed using a spin-echo sequence:
if Q2 is flipped midway through the free-evolution window,
its $\sigma_z$ expectation changes sign, effectively reversing the sign of
the accumulated ZZ phase.
By comparing the echo signal with and without the spectator flip,
one isolates the conditional phase
\begin{equation}
\Phi_{ZZ} = \int_0^\tau \frac{\zeta}{\hbar}\,s_2(t)\,dt,
\end{equation}
where $s_2(t)=\pm1$ tracks the $z$-projection of Q2.
This time-domain method thus provides a calibration of $\zeta$ without
requiring frequency-domain spectroscopy: the conditional phase
between the two Ramsey/echo traces grows linearly in time at a rate of
$\zeta/\hbar$.

In practice, plotting the Ramsey (or echo) phase of one qubit as a function
of the other qubit’s state or drive phase yields two oscillations whose relative
phase slope is $\zeta/\hbar$.
This direct phase accumulation measurement is commonly used to calibrate
the static ZZ coupling or to verify its suppression via echo-like
decoupling or bias tuning. In the present manuscript, we have used a modified but equivalent version of this experiment.

\subsection{Algorithm for maximizing the ZZ interaction}
The static ZZ interaction produces an energy shift of the doubly-excited computational state $\ket{11}$ relative to the sum of the single-excitation energies. We quantify its strength as
\begin{equation}
\zeta \;=\; E_{\ket{11}} - E_{\ket{10}} - E_{\ket{01}} + E_{\ket{00}},
\label{eq:zz_strength}
\end{equation}
where $E_{\ket{\gamma}}$ denotes the eigenenergy adiabatically connected to the bare state $\ket{\gamma}$.

Because $\zeta$ originates from hybridization of $\ket{11}$ with non-computational two-excitation states (e.g., $\ket{20}$ and $\ket{02}$), it is not captured by a strict two-level truncation of each qubit. We therefore compute $\zeta$ from a multi-level Hamiltonian obtained via black-box quantization and truncated to a fixed maximum excitation number.

We optimize (seek) circuit parameters $\mathbf{x}$ (capacitances and Josephson energies) that maximize $\zeta(\mathbf{x})$ subject to fabrication and operating constraints. For each candidate $\mathbf{x}$, we: (i) construct the black-box Hamiltonian $\hat{H}(\mathbf{x})$ obtained via \texttt{QuCat} \cite{GelySteele2020QuCAT}, (ii) truncate the Hilbert space to $N_{\mathrm{exc}}$ total excitations in a Fock basis, (iii) diagonalize to obtain $E_{\ket{00}},E_{\ket{01}},E_{\ket{10}},E_{\ket{11}}$, and (iv) evaluate $\zeta$ using Eq.~\eqref{eq:zz_strength}. The constraints enforce: (C1) the qubit frequencies within a target band, (C2) a minimum for the anharmonicity, (C3) bounds on all capacitances, (C4) a minimum for the transmon ratio $E_J/E_C$, and (C5) the operation in the dispersive regime via a threshold on $J/\Delta$.

To solve this constrained, potentially non-smooth optimization, we use the differential evolution method as implemented in \texttt{SciPy} \cite{Virtanen2020}, following Storn and Price \cite{Storn1997}. The algorithm, shown below in Alg. \ref{alg:algorithmopt}, is gradient-free and explores the design space by evolving a population of candidate circuits; constraint-handling is enforced either by the explicit rejection of infeasible candidates or by a penalty-augmented fitness. The output is an optimized design $\mathbf{x}^\star$ that satisfies all constraints while maximizing $\zeta$.

\begin{figure}
\begin{center}
\fbox{%
\begin{minipage}{0.95\linewidth}
\textbf{Algorithm 1. Constrained maximization of the static ZZ interaction (differential evolution)}\\[4pt]
\textbf{Input:} design variables $\mathbf{x}$ (capacitances, $E_J$) with bounds; truncation $N_{\mathrm{exc}}$; constraint thresholds (C1)--(C5); population size $N_p$; generations $G$.\\
\textbf{Output:} $\mathbf{x}^\star$ maximizing $\zeta$.\\[4pt]

\textbf{Procedure}
\begin{enumerate}
\item Initialize a population $\{\mathbf{x}_k\}_{k=1}^{N_p}$ by sampling uniformly within the parameter bounds.
\item For $g=1,\dots,G$:
  \begin{enumerate}
  \item For $k=1,\dots,N_p$:
    \begin{enumerate}
    \item Generate a trial vector $\tilde{\mathbf{x}}_k$ using differential-evolution mutation and crossover.
    \item Construct the black-box Hamiltonian $\hat{H}(\tilde{\mathbf{x}}_k)$ and truncate to $N_{\mathrm{exc}}$ total excitations.
    \item Diagonalize $\hat{H}(\tilde{\mathbf{x}}_k)$ and extract $E_{\ket{00}},E_{\ket{01}},E_{\ket{10}},E_{\ket{11}}$.
    \item Compute $\zeta(\tilde{\mathbf{x}}_k)$ using Eq.~\eqref{eq:zz_strength}.
    \item If $\tilde{\mathbf{x}}_k$ satisfies (C1)--(C5) and $\zeta(\tilde{\mathbf{x}}_k)\ge \zeta(\mathbf{x}_k)$, accept $\mathbf{x}_k \leftarrow \tilde{\mathbf{x}}_k$; otherwise reject and keep $\mathbf{x}_k$.
    \end{enumerate}
  \end{enumerate}
\item Return $\mathbf{x}^\star=\arg\max_{\mathbf{x}_k}\zeta(\mathbf{x}_k)$.
\end{enumerate}
\end{minipage}}
\end{center}
\label{alg:algorithmopt}
\end{figure}

\subsection{Dynamical blockade simulations}
\subsubsection*{Effective blockade Hamiltonian}
\label{app:effblockade}

The black-box quantization procedure provides the parameters of an effective two-qubit model: namely, the qubit transition frequencies $\omega_i$, the static ZZ interaction strength $\zeta$, and (optionally) residual transverse couplings. In this section, we use these parameters to build a minimal time-dependent Hamiltonian describing the \emph{dynamical blockade}: by driving both qubits on their dressed single-excitation transitions, population is transferred within the $\{\ket{00},\ket{01},\ket{10}\}$ manifold while the doubly-excited state $\ket{11}$ is energetically shifted by $\zeta$ and thus remains off-resonant when $|\zeta|\gg \Omega_i$.

To avoid confusion with the physical time-evolution propagator, we denote the (time-dependent) change-of-frame operator by $\hat{U}_{\mathrm{frame}}(t)$ and define
\begin{equation}
\ket{\psi_{\mathrm{rot}}(t)}=\hat{U}_{\mathrm{frame}}(t)\ket{\psi_{\mathrm{lab}}(t)},
\end{equation}
with
\begin{equation}
\hat{U}_{\mathrm{frame}}(t)=
\exp\!\left(+i\omega_{d,1}t\,\frac{\sigma_z^{(1)}}{2}\right)
\otimes
\exp\!\left(+i\omega_{d,2}t\,\frac{\sigma_z^{(2)}}{2}\right).
\label{eq:Uframe_def}
\end{equation}
Using this convention, the rotating-frame Hamiltonian can be expressed as
\begin{equation}
\hat{H}_{\mathrm{rot}}(t)=
\hat{U}_{\mathrm{frame}}(t)\,\hat{H}_{\mathrm{lab}}(t)\,\hat{U}_{\mathrm{frame}}^\dagger(t)
-i\hbar\,\hat{U}_{\mathrm{frame}}(t)\frac{\partial \hat{U}_{\mathrm{frame}}^\dagger(t)}{\partial t}.
\label{eq:Hrot_def}
\end{equation}

We model the lab-frame Hamiltonian as
\begin{equation}
\hat{H}_{\mathrm{lab}}(t)=
\sum_{i=1}^2 \frac{\hbar\omega_i}{2}\sigma_z^{(i)}
+\frac{\hbar\zeta}{4}\sigma_z^{(1)}\sigma_z^{(2)}
+\sum_{i=1}^2 \hat{H}_{\mathrm{d},i}(t)
+\hat{H}_{\mathrm{bb}},
\end{equation}
where $\omega_i$ and $\zeta$ are taken from the black-box quantization results. Since $[\sigma_z^{(i)},\hat{U}_{\mathrm{frame}}]=0$, the diagonal terms are unchanged by conjugation. The frame-derivative term gives
\begin{equation}
-i\hbar\,\hat{U}_{\mathrm{frame}}\frac{\partial \hat{U}_{\mathrm{frame}}^\dagger}{\partial t}
=
-\sum_{i=1}^2 \frac{\hbar\omega_{d,i}}{2}\sigma_z^{(i)},
\end{equation}
so that the rotating-frame Hamiltonian contains the detunings
\begin{equation}
\hat{H}_{\mathrm{rot}} \supset
\sum_{i=1}^2 \frac{\hbar\Delta_i}{2}\sigma_z^{(i)}
+\frac{\hbar\zeta}{4}\sigma_z^{(1)}\sigma_z^{(2)},
\qquad
\Delta_i\equiv \omega_i-\omega_{d,i}.
\label{eq:detunings}
\end{equation}

We take a single-qubit resonant drive in the lab frame
\begin{equation}
\hat{H}_{\mathrm{d},i}(t)=\hbar\Omega_i(t)\sin(\omega_{d,i}t)\,\sigma_y^{(i)},
\label{eq:drive_lab}
\end{equation}
where $\Omega_i(t)$ is the (possibly shaped) Rabi rate. Using
\begin{equation}
e^{i\gamma\sigma_z}\sigma_y e^{-i\gamma\sigma_z}
=\cos(2\gamma)\,\sigma_y+\sin(2\gamma)\,\sigma_x,
\end{equation}
we obtain
\begin{equation}
\hat{U}_{\mathrm{frame},i}\,\sigma_y^{(i)}\,\hat{U}_{\mathrm{frame},i}^\dagger
=
\cos(\omega_{d,i}t)\,\sigma_y^{(i)}
+\sin(\omega_{d,i}t)\,\sigma_x^{(i)}.
\label{eq:rot_y}
\end{equation}
Substituting Eq.~\eqref{eq:rot_y} into Eq.~\eqref{eq:drive_lab} gives a constant term plus and terms oscillating at $2\omega_{d,i}$. Under the rotating-wave approximation (dropping $2\omega_{d,i}$ terms),
\begin{equation}
\hat{U}_{\mathrm{frame}}\,\hat{H}_{\mathrm{d},i}(t)\,\hat{U}_{\mathrm{frame}}^\dagger
\;\xrightarrow{\mathrm{RWA}}\;
\frac{\hbar\Omega_i(t)}{2}\,\sigma_x^{(i)}.
\label{eq:drive_rwa}
\end{equation}
(Equivalently, starting from a $\cos(\omega_{d,i}t)\sigma_x^{(i)}$ lab drive yields the same RWA Hamiltonian up to a transverse phase convention.)

Collecting Eqs.~\eqref{eq:detunings} and \eqref{eq:drive_rwa}, we arrive at
\begin{equation}
\hat{H}_{\mathrm{rot}}(t)=
\sum_{i=1}^2 \frac{\hbar\Delta_i}{2}\sigma_z^{(i)}
+\frac{\hbar\zeta}{4}\sigma_z^{(1)}\sigma_z^{(2)}
+\sum_{i=1}^2 \frac{\hbar\Omega_i(t)}{2}\sigma_x^{(i)}
+\hat{H}_{\mathrm{bb,rot}}(t).
\label{eq:Hrot_full}
\end{equation}

Using the identity
\begin{equation}
\frac{\zeta}{4}\sigma_z^{(1)}\sigma_z^{(2)}
=
\zeta\,\ket{11}\!\bra{11}
-\frac{\zeta}{4}I
+\frac{\zeta}{4}\big(\sigma_z^{(1)}+\sigma_z^{(2)}\big),
\label{eq:zz_projector}
\end{equation}
and choosing the drive frequencies
\begin{equation}
\omega_{d,i}=\omega_i+\frac{\zeta}{2}
\qquad\Rightarrow\qquad
\Delta_i=-\frac{\zeta}{2},
\end{equation}
the single-qubit $Z$-terms cancel exactly:
\begin{equation}
\frac{\hbar\Delta_i}{2}\sigma_z^{(i)}+\frac{\hbar\zeta}{4}\sigma_z^{(i)}=0.
\end{equation}
Dropping the global energy shift term $-\hbar\zeta I/4$, the effective Hamiltonian used in simulations becomes
\begin{equation}
\hat{H}_{\mathrm{eff}}(t)=
\hbar\zeta\,\ket{11}\!\bra{11}
+\sum_{i=1}^2 \frac{\hbar\Omega_i(t)}{2}\sigma_x^{(i)}.
\label{eq:H_eff}
\end{equation}
This explicitly shows the blockade mechanism: $\ket{11}$ is detuned by $\hbar\zeta$ from the driven manifold, and is equivalent to the Hamiltonian we reported in the main text.

To assess the effect of non-ZZ interactions, we optionally include
\begin{equation}
\hat{H}_{\mathrm{bb}}=
\hbar J_{xx}\,\sigma_x^{(1)}\sigma_x^{(2)}+
\hbar J_{yy}\,\sigma_y^{(1)}\sigma_y^{(2)}.
\end{equation}
In the rotating frame, $\hat{H}_{\mathrm{bb,rot}}(t)$ contains components oscillating at the sum and difference frequencies
\begin{equation}
\omega_\Sigma=\omega_{d,1}+\omega_{d,2},
\qquad
\omega_\Delta=\omega_{d,1}-\omega_{d,2}\approx \omega_1-\omega_2.
\end{equation}
The $\omega_\Sigma$ terms are rotating fast and are therefore neglected under the RWA. For non-degenerate qubits satisfying
\begin{equation}
|\omega_\Delta|\gg \max_i \Omega_i,\;|\zeta|,\;|J_{xx}|,\;|J_{yy}|,
\end{equation}
the remaining $\omega_\Delta$ terms also average out on the timescale of the driven dynamics, so that $\hat{H}_{\mathrm{bb,rot}}(t)\approx 0$ and Eq.~\eqref{eq:H_eff} provides an accurate description of the dynamical blockade simulations.
The Hamiltonian above is the one used in the main text for fitting the data. Note that given the capacitances and the characteristics of the Josephson junctions, this is a zero parameter fit. The results are shown below in Fig.~\ref{fig:ryddyn}.

\begin{figure}[H]
    \centering
    \includegraphics[width=1.0\linewidth]{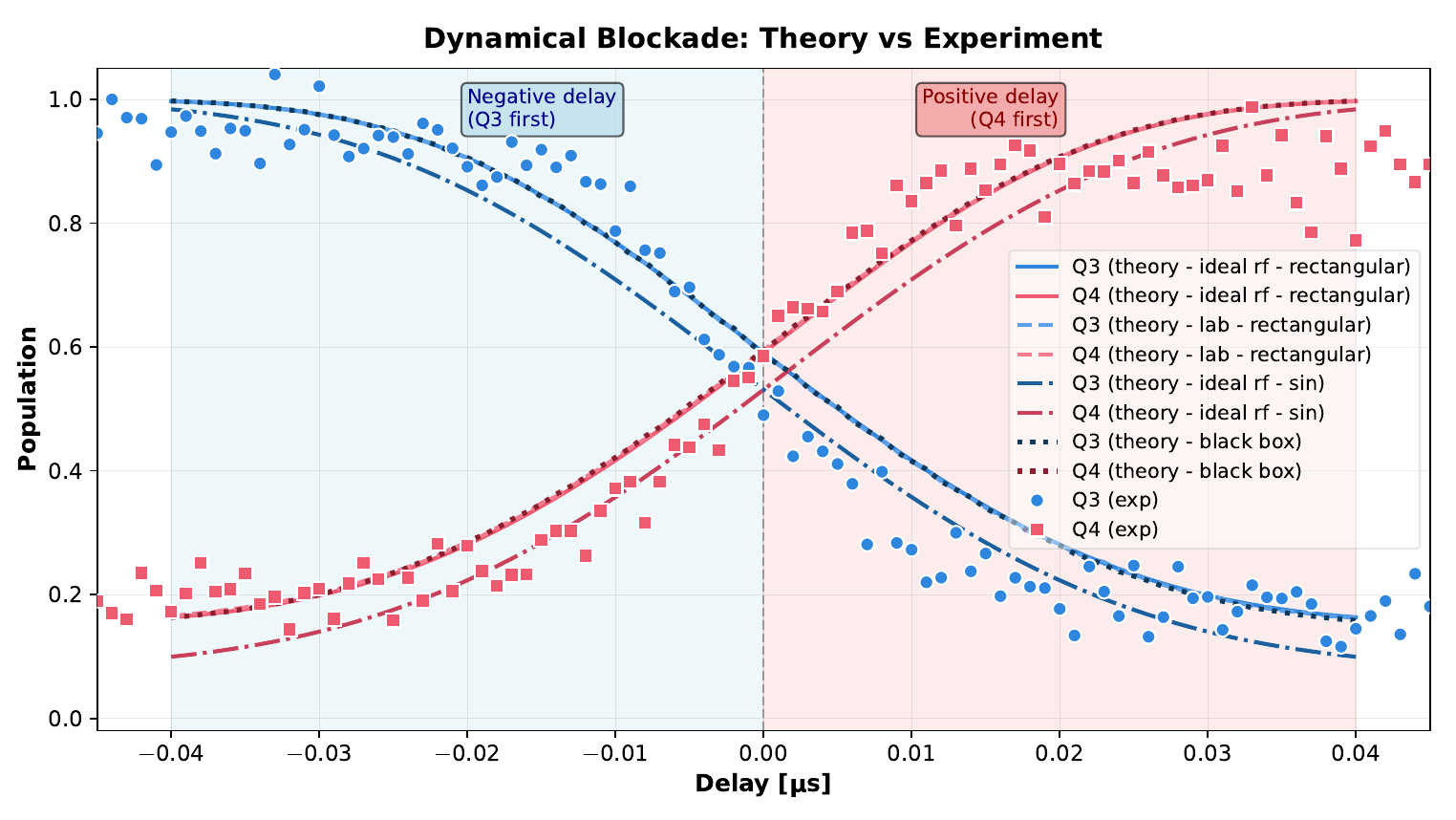}
    \caption{Qubit populations in the dynamical Blockade regime: experimental fit under various considerations and protocols. The dots are the experimental values, while the solid and dashed curves represent models. The notation is the following: rectangular vs sin implies a bang bang or the realistic pulses sent to the chip to show the difference between the two protocols. Black box implies that we are using the full black box Hamiltonian, not neglecting the XX+YY interactions. Ideal vs Lab implies in the full Hamiltonian vs the Lab frame.}
    \label{fig:ryddyn}
\end{figure}

\subsubsection*{Simulations}
From the circuit quantization made via the blackbox quantization technique in the previous section, we are able to write down the truncated static Hamiltonian:
\begin{equation}
    \hat{H}_0 := \beta_0 {\bm 1}^{(1)} \otimes {\bm 1}^{(2)} +  \beta_1 {\bm 1}^{(1)} \otimes \sigma_z^{(2)} +  \beta_2 \sigma_x^{(1)} \otimes \sigma_x^{(2)} +  \beta_3 \sigma_y^{(1)} \otimes \sigma_y^{(2)} +  \beta_4 \sigma_z^{(1)} \otimes {\bm 1}^{(2)} +  \beta_5 \sigma_z^{(1)} \otimes \sigma_z^{(2)} \ ,
\end{equation}
where the coefficients $\beta_i$ are given. In our specific case, they are reported in Table~\ref{tab:coeffs}.

\begin{table}[h!]
\centering
\begin{tabular}{c c c }
\hline
Coefficient & Value [Hz] for $Q_1$ and $Q_2$  & Value [Hz] for $Q_3$ and $Q_4$ \\
\hline
$\beta_0$  & $-13.72426610\times10^{9}$  & $6.57831254\times10^{9}$ \\
$\beta_1$  & $3.18923396\times10^{9}$ & $3.94913132\times10^{9}$ \\
$\beta_2$  & $8.05325187\times10^{6}$ & $5.34625496\times10^{6}$  \\
$\beta_3$  & $1.69065442\times10^{6}$ & $ 2.31009032\times10^{6}$  \\
$\beta_4$  & $4.59299123\times10^{9}$ & $2.77908918\times10^{9}$  \\
$\beta_5$  & $3.81259047\times10^{6}$ & $5.07031322\times10^{6}$ \\
\hline
\end{tabular}
\caption{Coefficients $\beta_i$ used in the model for both the qubits $Q_1$ and  $Q_2$ (in the main text) and the qubits $Q_3$ and  $Q_4$ for Figs.\ref{fig:ryddyn}-\ref{fig:qutip-sim}}
\label{tab:coeffs}
\end{table}

From Eq.~\eqref{eq:staticHsi}, we can identify the qubit frequencies as $\omega_1 := 2 \beta_4$, $\omega_2 := 2 \beta_2$, and the ZZ interaction strength as $\zeta := 4 \beta_5$.

To reproduce the blockade experiment, we need to include the $\pi$-pulse control as two driving Hamiltonian terms, one for each qubit:
\begin{equation}
    \hat{H}_{{\rm drive}, i} := \Omega_i(t) \sin(\omega_{{\rm d},i} t + \phi_i(t)) \sigma_{z}^{(i)} \, .
\end{equation}
The total Hamiltonian of the two-qubit system is therefore
\begin{equation}
    \hat{H}_{\rm tot} := \hat{H}_0 + \sum_{i=1,2} \hat{H}_{{\rm drive},i} \, .
\end{equation}
In the case of purely ZZ-interacting qubits ($\beta_0 = \beta_2 = \beta_3 = 0$), the energy difference between the first excited states, $|10\rangle_{12}$ and $01\rangle_{12}$, and the ground state $|00\rangle_{12}$ can be obtained from the spectrum of the static Hamiltonian, yielding
\begin{eqnarray}
    &&\Delta E_1 := E_{10} - E_{00} = \omega_1 - \zeta/2 \ ,\\
    &&\Delta E_2 := E_{01} - E_{00} = \omega_2 - \zeta/2\ ,
\end{eqnarray}
in units of $\hbar$. Therefore, to perform a $\pi$-pulse on the first qubit, we must drive it at the resonant frequency $\omega_{\rm d,1} = \Delta E_1$, inducing the transition $|00\rangle \to |10\rangle$, while to induce the transition $|00\rangle \to |01\rangle$, we have to drive the second qubit at frequency $\omega_{\rm d,2} = \Delta E_2$. Defining the delay time $\tau$ as the time difference between the $\pi$-pulse on the first and second qubit, we are able to test the blockade regime. 

\begin{figure}[H]
    \centering
    \includegraphics[width=1.0\linewidth]{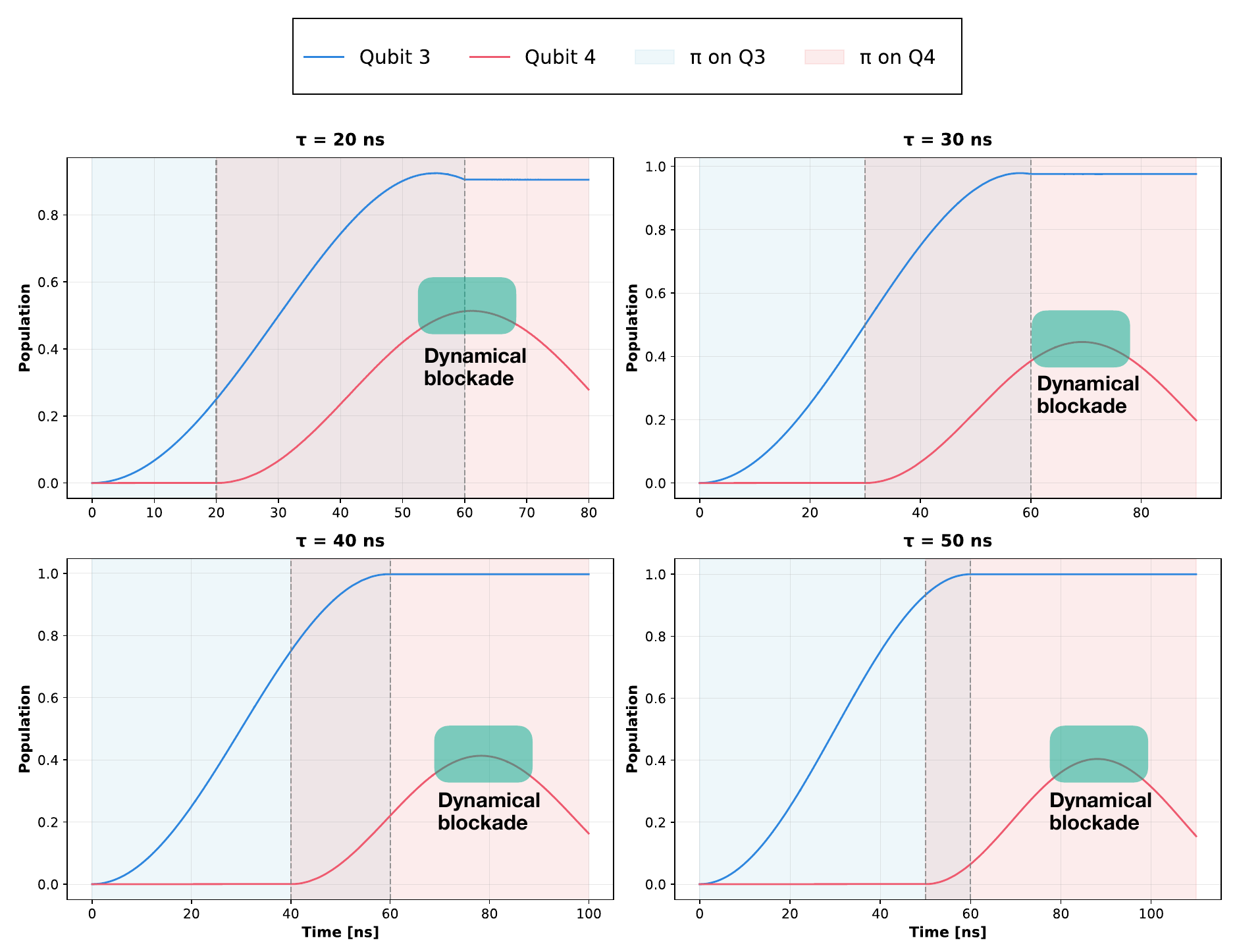}
    \caption{Dynamical blockade simulation experiment: population versus time for positive delay times $\tau = \{20,30,40,50\}$ ns. The Hamiltonian model used has no RWA. The population inversion is due to the inability of the second qubit to get excited.}
    \label{fig:qutip-sim}
\end{figure}

In Fig.~\ref{fig:qutip-sim}, we show a simulation neglecting approximations (such as the RWA), where we plot the population vs. time. It is evident that the dynamical blockade effect is induced by the strong ZZ interaction.

\end{document}